\documentclass[preprint2]{aastex}

\usepackage{graphicx}
\usepackage{psfig}
\usepackage{epsfig}
\usepackage{amsmath}

\newcommand{\msun}{\,\hbox{$M_{\odot}$}}

\newcommand{\kms}{\,\hbox{\hbox{km}\,\hbox{s}$^{-1}$}}
\newcommand{\mbh}{\,\hbox{$M_{\rm BH}$}}
\newcommand{\msigma}{\,\hbox{$M_{\rm BH}-\sigma$}}

\newcommand{\ha}{\,\hbox{$H_{\rm \alpha}$}}
\newcommand{\hb}{\,\hbox{$H_{\rm \beta}$}}

\newcommand{\spi}{{\it Spitzer}}
\newcommand{\ang}{\,\hbox{\AA}}
\newcommand{\nii}{\,\hbox{[\ion{N}{2}]}}
\newcommand{\neii}{\,\hbox{[\ion{Ne}{2}]}}
\newcommand{\neiii}{\,\hbox{[\ion{Ne}{3}]}}
\newcommand{\nev}{\,\hbox{[\ion{Ne}{5}]}}
\newcommand{\siv}{\,\hbox{[\ion{S}{4}]}}
\newcommand{\oiv}{\,\hbox{[\ion{O}{4}]}}
\newcommand{\oi}{\,\hbox{[\ion{O}{1}]}}
\newcommand{\oiii}{\,\hbox{[\ion{O}{3}]}}

\newcommand{\feii}{\,\hbox{[\ion{Fe}{2}]}}
\newcommand{\sstar}{\,\hbox{$\sigma_{*}$}}

\newcommand{\sneii}{\,\hbox{$\sigma_{\rm [Ne~\sc{II}]}$}}
\newcommand{\ssiv}{\,\hbox{$\sigma_{\rm [S~\sc{IV}]}$}}
\newcommand{\sneiii}{\,\hbox{$\sigma_{\rm [Ne~\sc{III}]}$}}
\newcommand{\snev}{\,\hbox{$\sigma_{\rm [Ne~\sc{V}]}$}}
\newcommand{\soiv}{\,\hbox{$\sigma_{\rm [O~\sc{IV}]}$}}
\newcommand{\soiii}{\,\hbox{$\sigma_{\rm [O~\sc{III}]}$}}

\newcommand{\um}{\,\hbox{$\mu$m}}

\shorttitle{ Resolved narrow lines in the IR }

\shortauthors{Dasyra et al.}

\begin{document}

\title{A view of the narrow-line region in the infrared: active galactic nuclei with resolved fine-structure lines in the \spi\ archive}

\author{K. M. Dasyra\altaffilmark{1,2},  L. C. Ho\altaffilmark{3}, H. Netzer\altaffilmark{4}, F. Combes\altaffilmark{2}, B. Trakhtenbrot\altaffilmark{4}, E. Sturm\altaffilmark{5}, 
L. Armus\altaffilmark{6}, D. Elbaz\altaffilmark{1}
}

\altaffiltext{1}{Laboratoire AIM, CEA/DSM-CNRS-Universit\'e Paris Diderot, Irfu/Service
dÕAstrophysique, CEA Saclay, F-91191 Gif-sur-Yvette, France}
\altaffiltext{2}{Observatoire de Paris, LERMA (CNRS:UMR8112), 61 Av. de l\'\ Observatoire, F-75014, 
Paris, France }
\altaffiltext{3}{The Observatories of the Carnegie Institution for Science,
813 Santa Barbara St., Pasadena, CA 91101}
\altaffiltext{4}{School of Physics and Astronomy and the Wise Observatory, 
Raymond and Beverly Sackler Faculty of Exact Sciences, Tel-Aviv University, 
Tel-Aviv, 69978, Israel}
\altaffiltext{5}{Max-Planck-Institut f\"ur Extraterrestrische Physik, Postfach 1312, 85741 Garching, Germany}
\altaffiltext{6}{Spitzer Science Center, California Institute of Technology,
Mail Code 220-6, 1200 East California Blvd, Pasadena, CA 91125}

\begin{abstract}
We queried the \spi\ archive for high-resolution observations with the Infrared Spectrograph of optically selected active galactic nuclei 
(AGN) for the purpose of identifying sources with resolved  fine-structure lines that would enable studies of the narrow-line region (NLR) 
at mid-infrared wavelengths. By combining 298 \spi\ spectra with 6 {\it Infrared Space Observatory} spectra, we present kinematic information 
of the NLR for 81 z$\lesssim$0.3 AGN. We used the \nev , \oiv , \neiii , and \siv\ lines, whose fluxes correlate well with each other, to 
probe gas photoionized by the AGN. We found that the widths of the lines are, on average, increasing with the ionization potential of the 
species that emit them. No correlation of the line width with the critical density of the corresponding transition was found. The velocity 
dispersion of the gas, $\sigma$, is systematically higher than that of the stars, $\sigma _*$, in the AGN host galaxy, and it scales with 
the mass of the central black hole, \mbh . Further correlations between the line widths and luminosities $L$, and between $L$ and \mbh , are 
suggestive of a three dimensional plane connecting log(\mbh ) to a linear combination of log($\sigma$) and log($L$).  Such a plane can be 
understood within the context of gas motions that are driven by AGN feedback mechanisms, or virialized gas motions with a 
power-law dependence of the NLR radius on the AGN luminosity. The \mbh\ estimates obtained for 35 type 2 AGN from this plane are 
consistent with those obtained from the $\msigma _*$ relation.
\end{abstract}


\section{Introduction}
\label{sec:intro}

Meaningful statistical tests of whether the bulk of black hole (BH) growth precedes, is parallel, or follows the peak of star-formation activity need 
to rely on the comparison of several observable parameters as a function of look-back time. In addition to the comparison of the star-formation rate 
(SFR) with the BH accretion rate  (e.g., \citealt{marconi04}; \citealt{merloni04}), the comparison of the stellar mass with the BH mass, \mbh , that is 
already accumulated at any given redshift $z$ is also desirable. A significant impediment in performing the latter comparison comes from the fact that 
rapid BH growth often occurs in highly obscured environments, such as type 2 active galactic nuclei (AGN) and infrared (IR) bright galaxies. At low $z$
and at low (Seyfert-like) luminosities, $L$,  50$-$70\% of the AGN are of type 2 (\citealt{ho97b}; \citealt{schmitt01}; \citealt{hao05}).  At z$\sim$1 and at 
bolometric luminosities that exceed 10$^{12}$ $L_{sun}$, at least 1 out of every 3 AGN is thought to be of type 2 (\citealt{lacy07}; \citealt{gilli07}). 

In type 2 systems, the determination of \mbh\ is very challenging. Direct methods of measuring \mbh , e.g., by spatially resolved sub-parsec or parsec
scale kinematics of stars (\citealt{genzel97}), ionized gas (\citealt{harms94}; \citealt{maccheto97}), or water masers (\citealt{miyoshi95}), can only be applicable 
to very local objects.  An alternative method that is successfully applied to type 1 AGN even at high $z$ (\citealt{kaspi07}) uses the kinematics and the radial 
extent of gas clouds that are close enough to the BH to be gravitationally influenced by it. These clouds constitute the AGN broad line region (BLR). Emission 
lines tracing the BLR, such as H$\beta$, \ion{Mg}{2} and \ion{C}{4}, have typical full width at half maxima (FWHM) exceeding 2000 \kms . The BLR radius is determined by 
reverberation mapping experiments, which use the natural variability of the AGN continuum and the time-delayed response of the gas (de)excitation to compute 
the light travel time across the BLR (\citealt{blandford82}; \citealt{peterson93}; \citealt{kaspi00}). Assuming that the clouds are virialized, their distance from the 
BH and their velocity dispersion can be used to determine an enclosed dynamical mass and, thus,  \mbh . However, the method is not applicable to type 2 AGN 
where the BLR is obscured.
 
The lines that are seen in both type 1 and type 2 AGN are those that typically originate from gas clouds at a few pc  to a few hundreds of pc away from the 
BH (\citealt{netzer04}; \citealt{laor07} and references therein). These narrow-line-region (NLR) clouds have FWHM that are typically $\lesssim$1000 \kms . 
Using the \oiii\ 5007\ang\ line, \cite{nelson96} demonstrated that the bulk of the gas traced by \oiii\ has a velocity dispersion that is comparable to that of 
the stars in the host-galaxy bulge. Subsequently, \mbh\ was found to scale with the width of the \oiii\ emission (\citealt{nelson00}; \citealt{shields03}; 
\citealt{greene05}). A similar result was found for the \nii\ line at 6583\ang\ by \cite{ho09}. 

Optical NLR lines can suffer severe obscuration (\citealt{kauffmann03}). This makes IR wavelengths more reliable and often unique for 
investigating relations between NLR kinematics and \mbh . To optimize the use of narrow lines as gravitational potential tracers for obscured AGN, 
we performed a similar analysis in mid-infrared (MIR) wavelengths using \spi\ and {\it Infrared Space Observatory} (ISO) high-resolution spectra. We 
demonstrated that the widths of the NLR lines \nev\ at 14.32 \um\ and \oiv\ at 25.89 \um\ also scale with \mbh\ (\citealt{dasyra08}). This result could 
provide a means of probing NLR kinematics and weighing BH masses in obscured galaxies at high $z$ with the next generation IR telescopes. 

The goal of this paper is to perform an extended study of the NLR kinematics of optically selected type 1 and type 2 AGN in the MIR using all spectra that 
are available in the  \spi\  archive.  We aim to investigate for differences in the gas kinematics as traced by various fine-structure lines, to test how the gas 
velocities compare with \mbh , and to further estimate the masses of local obscured AGN. For all computations, we use a $\Lambda$CDM cosmology with 
$H_0$=70~km~s$^{-1}$~Mpc$^{-1}$, $\Omega_{m}$=0.3, and $\Omega_{\Lambda}$=0.7.


\section{The Sample}
\label{sec:sample}

To study the NLR gas kinematics as probed by MIR emission lines, we queried  the entire \spi\ archive for high-resolution observations
of AGN obtained with the Infrared Spectrograph (IRS; \citealt{houck04}). These spectra have a resolving power of $\sim$500 \kms , sufficient for 
resolving features in the NLR of several local AGN (\citealt{dasyra08}). We downloaded the reserved observations catalog (ROC) after the completion 
of \spi 's cryogenic mission to ensure that all archival data that can be used for this study are included in our sample. We found 1366 astronomical 
observation requests (AORs) containing high-resolution IRS spectra of extragalactic sources performed in either single-target or cluster mode. 

We matched all IRS targets with optical spectroscopic catalogs of AGN to identify type 1 sources with existing \mbh\ estimates and \oiii\ 5007\ang\  
detections that are required for the comparison of NLR kinematics at various wavelengths. For this purpose, we used i) the reverberation mapping 
catalogs of \cite{peterson04}, \cite{bentz06}, \cite{bentz09b}, and \cite{denney10}, ii) the spectroscopic subsample of the Palomar-Green (PG) QSOs 
(\citealt{boroson92}; \citealt{vestergaard06}), iii) the optical spectroscopic catalog of \cite{marziani03}, iv) the \cite{hokim09} catalog of type 1 AGN, v) the 
Sloan digital sky survey (SDSS) data release 4 AGN catalogs of \cite{kauffmann03}, of \cite{greene05}, and of \cite{netzer07}, and vii) the SDSS data release 
7 (DR7). To identify AGN in the DR7, we used the \cite{bpt} diagnostic diagram of \oiii /H$_{\beta}$ versus \nii /H$_{\alpha}$ with the AGN boundaries as updated 
by \cite{kewley06}. We also examined sources that have a QSO target type, or a broad-line or AGN spectral type subclass assigned by the SDSS pipeline.
These samples also include a few type 2 AGN, which we further complemented by matching all IRS targets with the \cite{mulchaey94}, \cite{turner97}, 
\cite{bassani99}, \cite{zakamska03},  \cite{reyes08}, \cite{bennert09}, and \cite{liu09} catalogs. A few additional optical spectra of type 2 AGN were found
in \cite{spinelli06}

\begin{figure}[t!]
\centering
\includegraphics[width=\columnwidth]{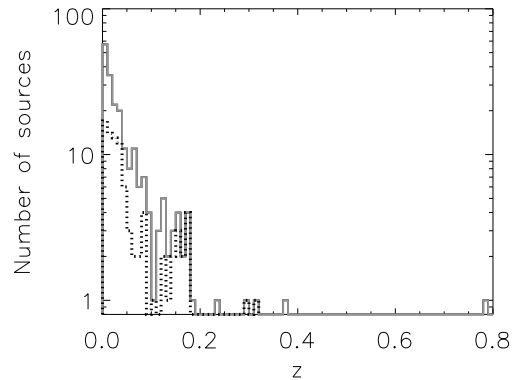}
\caption{\label{fig:z} {\footnotesize Redshift distribution of AGN with MIR \siv , \neiii , \oiv , or \nev\ detections (grey solid line),
and redshift distribution of AGN with resolved  \siv , \neiii , \oiv , or \nev\ lines (black dotted line).}}
\end{figure}

In total, we found and reduced datasets from 370 \spi\ AORs, corresponding to high-resolution spectra of 298 sources. Resolved NLR lines were determined 
from the inspection of the \siv\ 10.51\um  , \neiii\ 15.56\um , \oiv\ 25.89\um , or \nev\ 14.32\um\ profiles as described in Section~\ref{sec:reduction_line}. 
To these 298 AGN, we added 6 sources with fine structure lines that were similarly resolved by ISO short wavelength spectrograph 
(SWS) data (\citealt{sturm02}; \citealt{dasyra08}).  The redshift distribution of all 304 sources in the final sample is shown in Figure~\ref{fig:z}. The 42 type 1 AGN and 
the 39 type 2 AGN that have resolved MIR narrow lines are presented in Tables~\ref{tab:type1_sum} and~\ref{tab:type2_sum}, respectively.


\section{Data Reduction}
\label{sec:reduction}

\subsection{Spectral Extraction}
\label{sec:reduction_spec}
We downloaded the \spi\ S15.0 pipeline basic calibration data (BCD) files. We used both staring and mapping
observations, taken on either single-source or cluster mode. 

The BCD files of the long-high (LH) wavelength data of each source were processed by the IDL routine DARK SETTLE,
which is posted at the \spi\ Science Center (SSC) web site, to correct for gradations of the dark current along the detector that lead to order 
tilting and mismatch. Using the short-high (SH) BCD files and the dark-settle corrected LH BCD files, we computed the 
average frame for each set of on-source observations, namely, for each each target, nod position, and module. To identify 
cosmic ray hits, we also computed the median frame for the same set of observations and compared it to the average frame 
on a pixel by pixel basis. A pixel was flagged if the difference between the median and the average frame value differed by 
a factor of more than three times the root-mean square (rms) noise in the median frame. For this pixel, the average value 
was replaced by the median value. For
sources with available observations on sky\footnote{ Several programs whose scientific goals required only the use of line 
fluxes (but not the use of equivalent widths or continuum flux measurements) did not acquire sky observations given that 
the flux of a line does not depend on its underlying continuum level.}, we also computed the median sky frame, which we 
subtracted from the on-source frame.

The next step was the removal of bad and rogue (i.e., slowly varying time response) pixels. We used the individual sky frames 
to create a generic bad pixel mask for all sources  observed in a single \spi\ campaign, which we merged with the bad 
pixel map available at the SSC web site for the same campaign. We then merged this generic mask with the mask of 
each individual (on-source and on-sky) exposure to create the mask for each science frame, nod position, and module. 
We proceeded to further masking of outliers, i.e., pixels that were located a couple of columns away from the edges of 
each spectral order, and that had a value exceeding the rms noise of the science frame. Their values were replaced with 
the median value of the frame, computed using only pixels in the useful detector range. A final visual inspection and manual 
cleaning of the science frames was performed using the IRSCLEAN routine. All rogue pixels and outliers were flagged in the 
mask file. 

The uncertainty of the average on-source frame was calculated as the square root of the sum of the squares
of all individual uncertainty files, divided by the number of exposures. When sky observations were available,
we computed the sky frame uncertainty in a similar way and combined it with the on-source frame uncertainty
to produce the uncertainty of the final science frame. 

We used the science frame, together with its uncertainty and mask files as input to the SSC software SPICE, which produces 
a one-dimensional spectrum from a two-dimensional spectral image. To extract the spectra, we used the regular extraction 
mode, which equally weighs pixels when collapsing them along each row. We assumed a point source extraction, since the 
LH and SH apertures (22\farcs3$\times$11\farcs1 and 11\farcs3$\times$4\farcs7, respectively) are likely to include the bulk of 
the NLR emission for most sources in our sample. The end product of SPICE is the wavelength 
and flux calibrated spectrum for each individual order. 

We merged the spectra of the various orders to a single spectrum for each nod position, clipping noisy edges 
(between 2-25 pixels, depending on the order). This task was performed for both the SH and the LH datasets. 
The SH and LH spectra were then merged to produce the full-range spectrum per nod position. The final spectrum 
of each object was produced by averaging the one-dimensional, full-range spectra of the two nod
positions. Special care was taken not to merge sky-subtracted and non-sky-subtracted datasets for sources with 
observations from different programs. At the wavelengths of key atomic/molecular lines, only nod positions without bad 
pixels were used, when possible. For all other pixels with a bad pixel flag in one nod position, we only kept the value
of the second nod position when the two values differed by more than the local rms noise value.

\begin{figure}[ht!]
\centering
\includegraphics[width=6cm]{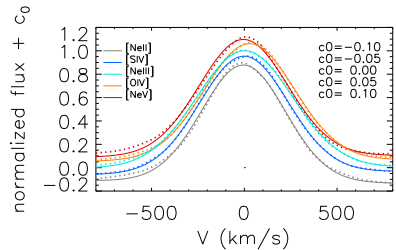}
\caption{\label{fig:systematic_offset}   \footnotesize{Stacked profiles (dotted lines) constructed using 29 $z$$<$0.02 individual sources with unresolved 
line detections at all wavelengths. Solid lines correspond to the Gaussian best-fit solutions of the stacked profiles. The differences in the line 
peak positions are indicative of wavelength calibration uncertainties.}}
\end{figure}

\begin{figure*}
\centering
\includegraphics[width=17cm]{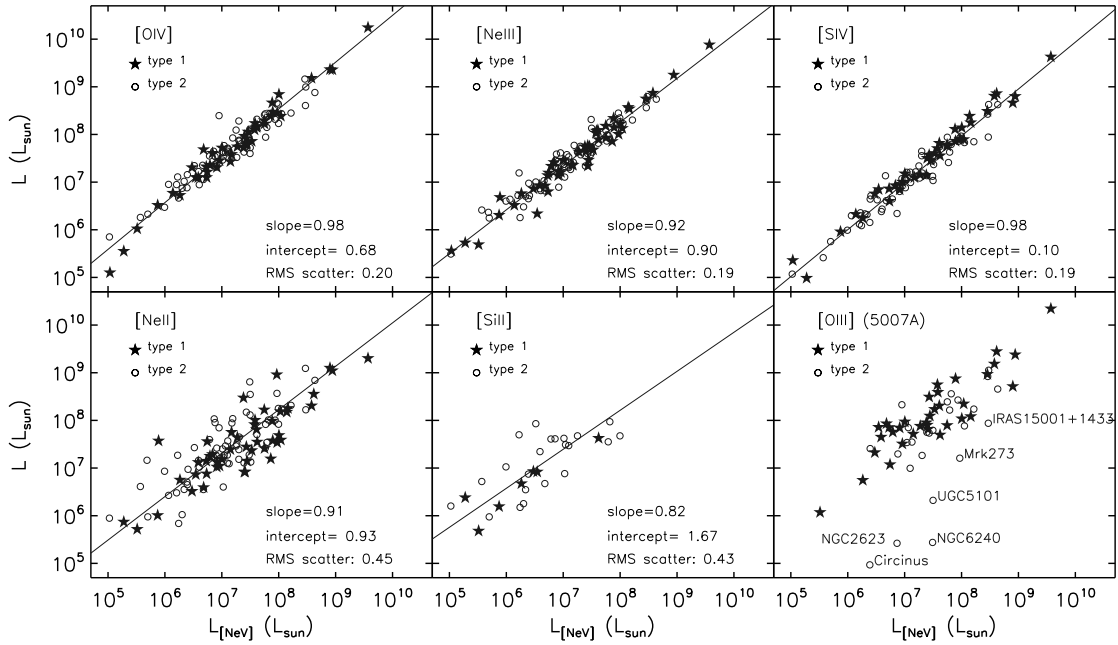}
\caption{\label{fig:line_luminosities}  \footnotesize{ Relation of the \nev\ luminosity, commonly used as an AGN indicator, with the luminosities of other 
fine-structure lines in MIR and optical wavelengths. The optical luminosities are not corrected for extinction.}}
\end{figure*}

\subsection{Line Fitting}
\label{sec:reduction_line}

As in \cite{dasyra08}\footnote{All previously presented data were reanalyzed to account for changes in the \spi\ pipeline, and 
for enhancements in our bad-pixel identification routines.}, we fitted all lines and their underlying continua using Gaussian 
and second-order polynomial functions, respectively. Gaussian fitting was preferred over direct FWHM computations which,
given the resolution of IRS, can only be reliably performed for few high-velocity systems (e.g., \citealt{spoon09b}). Gaussian 
fitting also suits best the studies of cloud kinematics on virial, regular, or symmetric motions, instead of clouds that are 
entrained by asymmetric outflows (\citealt{greene05}). 

To claim a line detection, we required that its signal-to-noise ratio (S/N) was greater than 3. 
To measure line widths, however, we only used lines with S/N$>$5 to avoid studying profiles of barely 
detected lines. We considered resolved all lines with $\sigma_m - \epsilon_m > \sigma_i+\epsilon_i$, where $\sigma_{m}$ is the 
measured velocity dispersion, $\sigma_i$ is the instrumental resolution  at a given wavelength divided by 2.35, and $\epsilon_i$ is 
the error of  $\sigma_i$. The average resolution value in the 12.0$-$18.0 \um\ range, which comprises all neon lines for the low 
$z$ galaxies, is 507$\pm$66 \kms. The 25.0$-$34.2 \um\ range, which comprises most of the \ion{O}{4} emission, has an
average resolution of 503$\pm$63 \kms.  The error of $\sigma_{m}$, $\epsilon_m$, encapsulates both measurement and 
instrumental uncertainties. It was computed as $(\epsilon_{st}^2+\epsilon_i^2)^{0.5}$, where $\epsilon_{st}$ is the standard 
deviation of the different velocity dispersion values that were obtained for each line when using different polynomial functions to
describe its underlying continuum. Intrinsic velocity dispersions, $\sigma$,  were computed as  $(\sigma_m^2-\sigma_i^2)^{0.5}$, converted 
to restframe, and presented in Tables~\ref{tab:type1_sum} and \ref{tab:type2_sum} for key fine-structure lines in all sources.

To assess possible systematic errors in the wavelength calibration, we normalized and stacked the line profiles of 29 sources
with detected but unresolved emission lines of \neii, \siv , \neiii, \oiv , and \nev\ (Figure~\ref{fig:systematic_offset}). The stacking of 
unresolved lines in different-$z$ sources permitted for a finer sampling of the instrumental resolution curve than that determined 
by the detector pixel size. We only used sources at $z$$<$0.02 to study systematic effects within a few pixels from the restframe 
wavelength of each line. This $z$ cutoff translates to  25 pixels at the center of the 
SH array range, 14.7 \micron . We found that line peak offsets due to such systematics are typically limited to 50 \kms .


\section{Results}
\label{sec:results}

\subsection{Identifying NLR Tracers in the MIR}
\label{sec:results_NLR_tracers}

Of the 304 AGN in our sample, 300 had spectral coverage at 10.51\um , and 143 showed \siv\ emission. Similarly, 226 sources had 
spectral coverage at 25.89\um , and 135 had \oiv\ detections. Thus, the typical line detection rate was of order $\sim$50\%.  

We plotted the luminosities of the most frequently observed MIR lines against that of \nev\ to identify fine-structure lines that can be reliably used as 
tracers of the gas that is photoionized by the AGN (Figure~\ref{fig:line_luminosities}). Given that photons of at least 97.12 eV are required to ionize \ion{Ne}{4} 
to \ion{Ne}{5}, its ionization source must be far-ultraviolet or soft X-ray radiation from the AGN. We find that the \nev\ luminosity is on a tight correlation 
of 0.2 dex scatter with the \oiv\ and \siv\ luminosities (Figure~\ref{fig:line_luminosities}; \citealt{pereira10}), and we further confirm that the \nev\ and \neiii\ 
luminosities trace each other well (\citealt{gorjian07}). Therefore, the \ion{Ne}{5}, \ion{O}{4}, \ion{Ne}{3}, and \ion{S}{4} ions are most likely to be primarily 
excited by the same mechanism. However, the scatter between the luminosities of \nev\ and \neii\ or [\ion{Si}{2}] is approximately twice as high, of 0.4$-$0.5 
dex. \ion{Ne}{2} and \ion{Si}{2} have low ionization potentials (21.56 and 8.15 eV, respectively), thus, a non-negligible part of their line emission could be 
tracing star-forming regions in the AGN host galaxy. 

The excellent, almost one-to-one correlation over several orders of magnitude in luminosity that is seen between the \nev\ 14.32\um\ and \siv\ 
10.51\um\ lines does not include any extinction correction. This result indicates that the bulk of the \siv\ emission is not strongly affected by the silicate 
absorption feature centered at 9.7 \um , which is seen in several of our targets with moderate or high optical depths, i.e. $\tau$$>$0.5. Thus, the geometric 
distribution of the silicates is likely to be more compact than the size of the NLR (see also \citealt{soifer02}; \citealt{tristram07}; \citealt{schweitzer08}). 
While obscuration effects do not significantly affect the relative fluxes of MIR lines, the \oiii\ 5007\ang\ emission can suffer from strong 
extinction primarily in type 2 AGN (Figure~\ref{fig:line_luminosities}).

\begin{figure*}
\centering
\includegraphics[width=15.8cm]{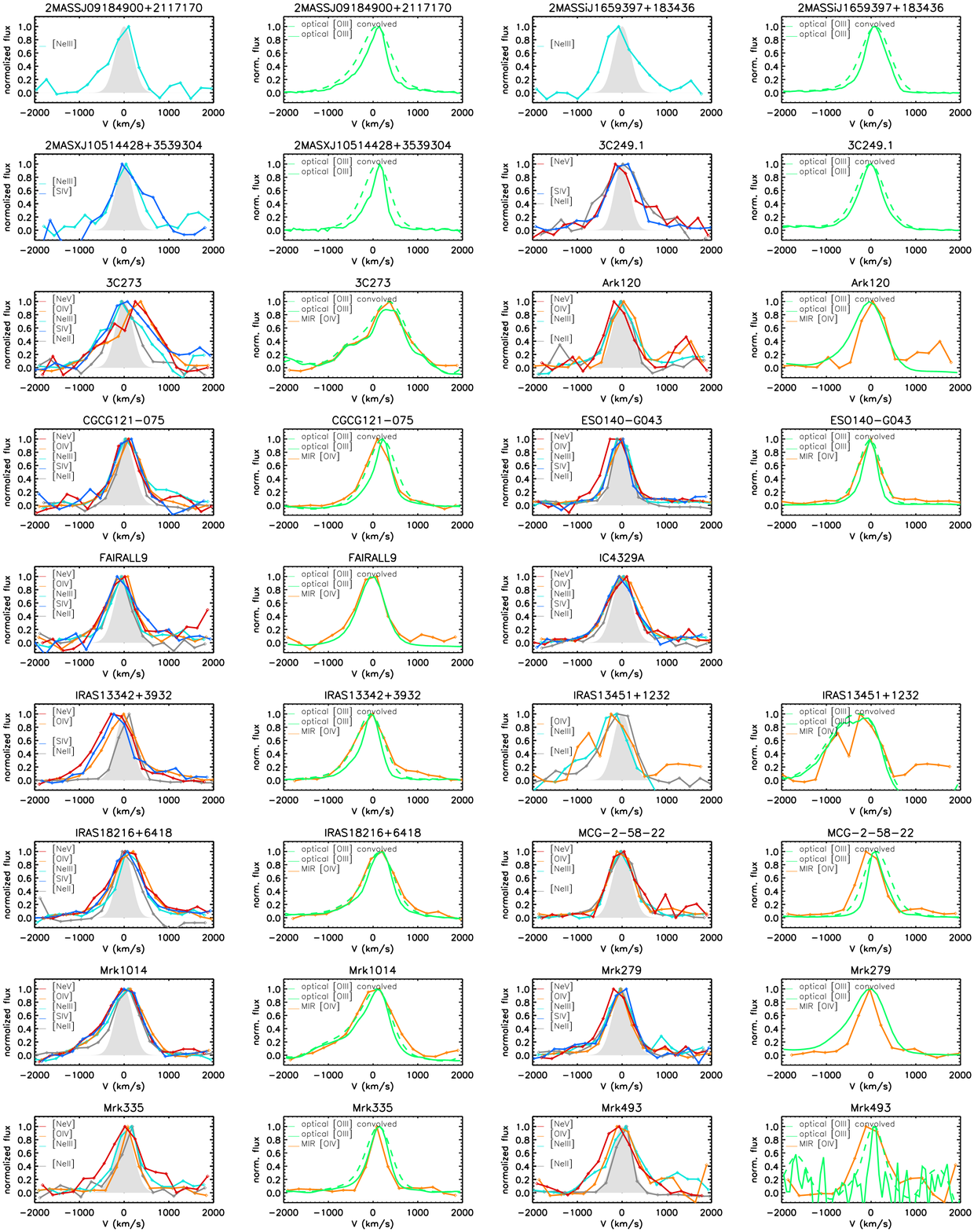}
\caption{\label{fig:profiles1}   \footnotesize{ Profiles of fine-structure lines in type 1 AGN with at least one resolved NLR line in the MIR. 
For each source, the left panel presents the profiles of the MIR lines, while the right panel shows the comparison of the \oiv\ 25.89 \um\  
line with the optical \oiii\ 5007\ang\ line, convolved to the resolution of the IRS data. The filled area corresponds to the average IRS 
resolution at the wavelengths of those lines.}}
\end{figure*}
\begin{figure*} \includegraphics[width=15.8cm]{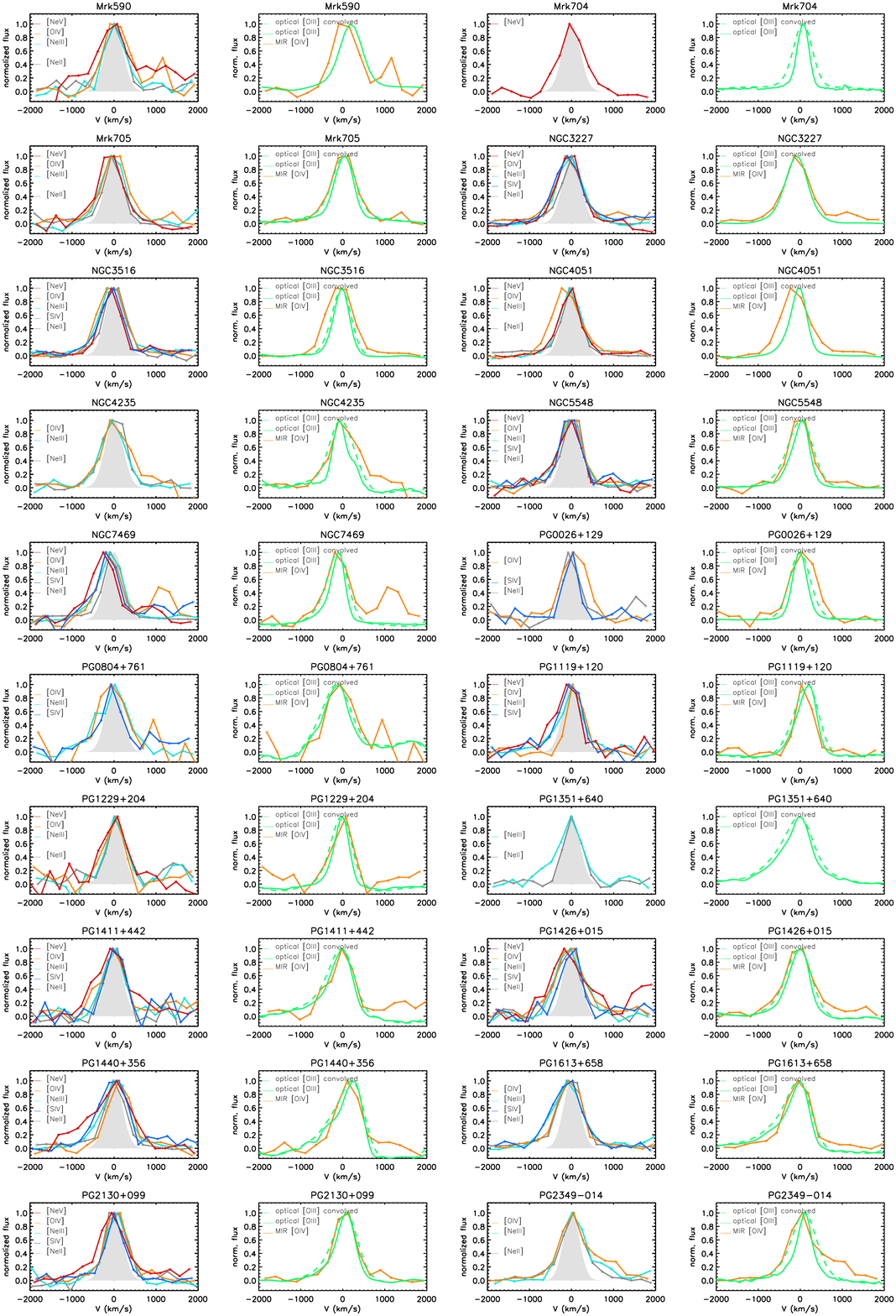} \newline
\footnotesize{  Figure~\ref{fig:profiles1}$-$ continued. }
 \end{figure*} 

\begin{figure*}
\centering
\includegraphics[width=15.8cm]{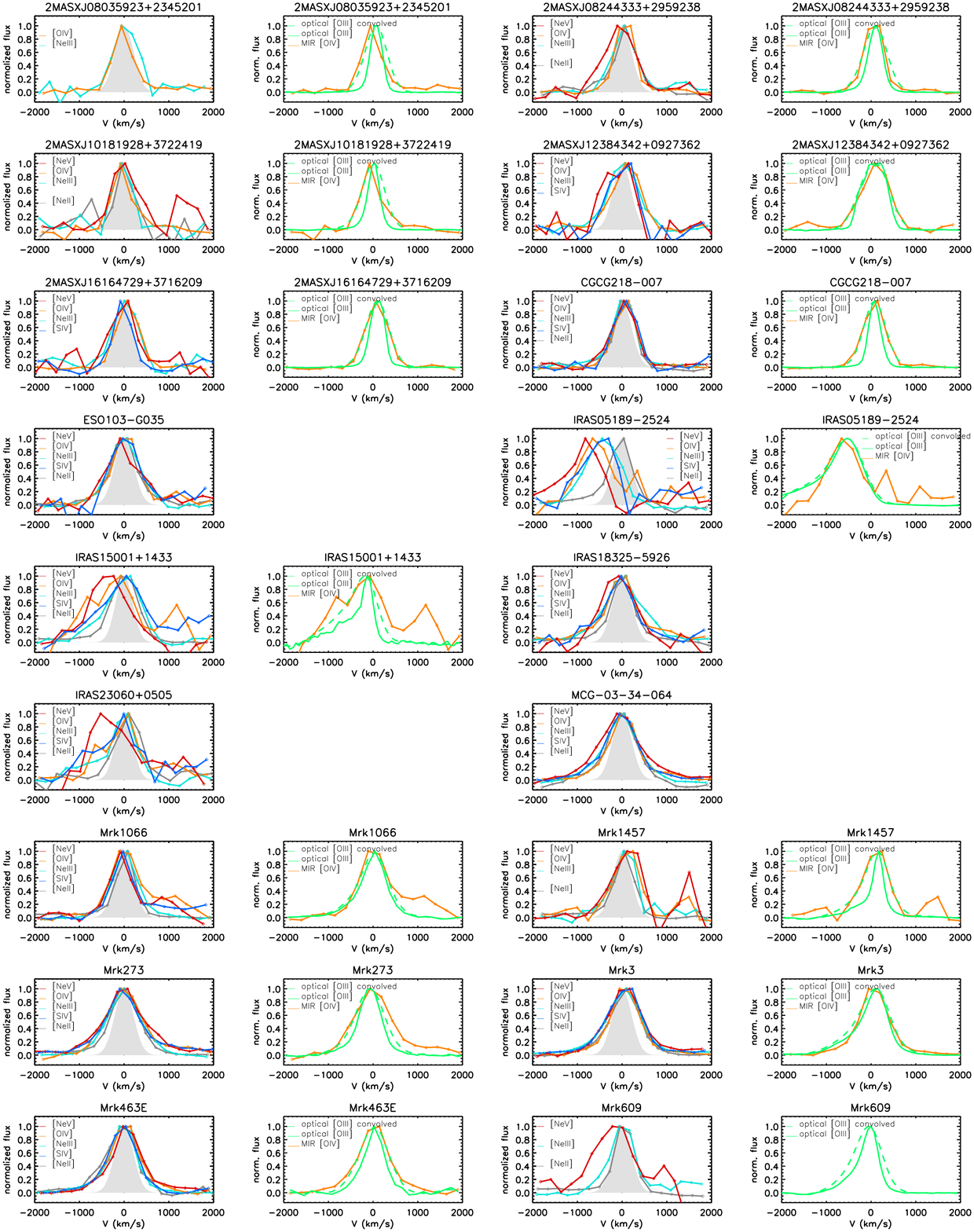}
\caption{\label{fig:profiles2}  \footnotesize{ Profiles of fine-structure lines in type 2 AGN with at least one NLR line resolved in the MIR.  }}
\end{figure*}
\begin{figure*} \includegraphics[width=15.8cm]{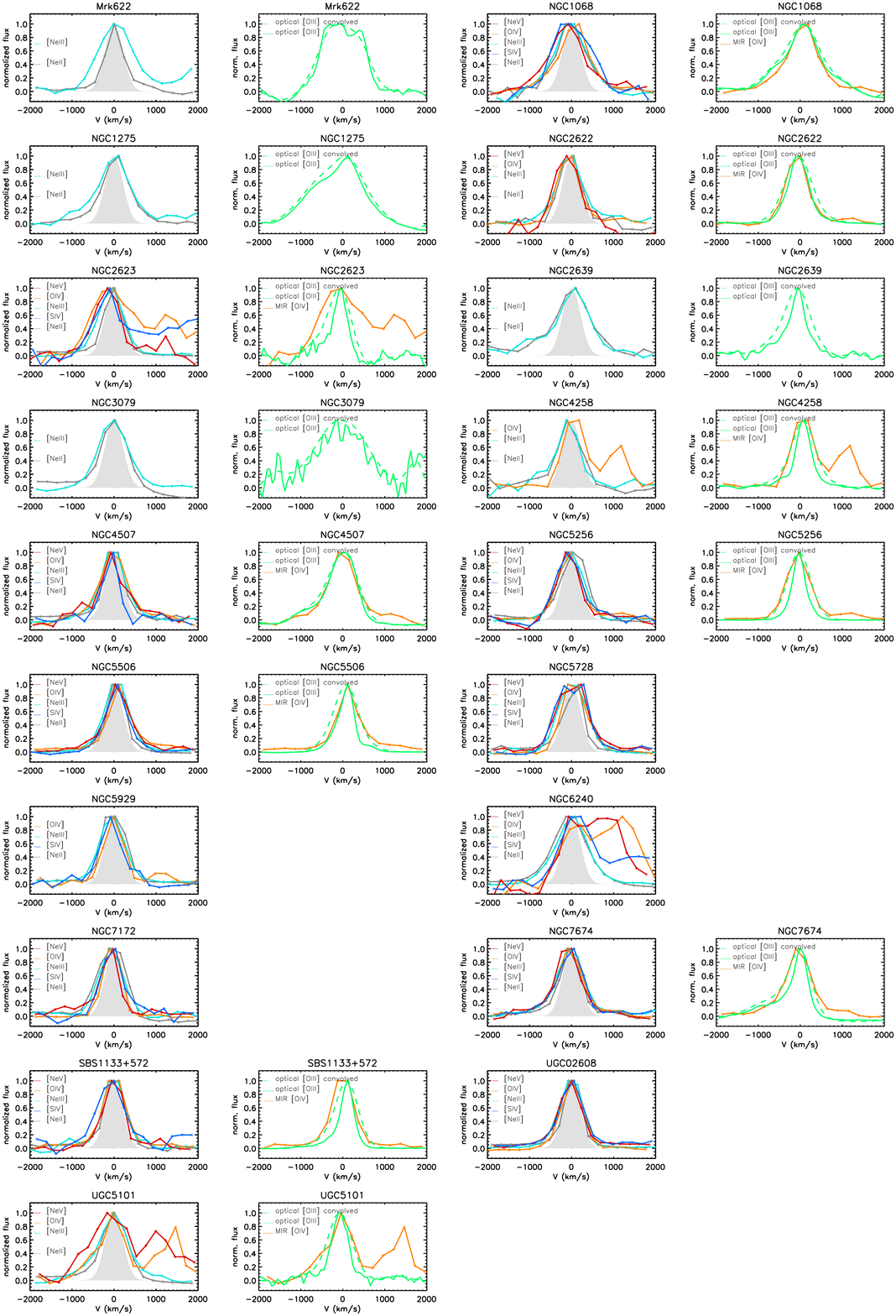}  \newline
\footnotesize{ Figure~\ref{fig:profiles2}$-$ continued. } 
\end{figure*} 

\subsection{Line Profiles: Velocity Dispersion of Different Ionized Gas Components}
\label{sec:results_widths}

\begin{figure}[h!]
\centering
\includegraphics[width=7cm]{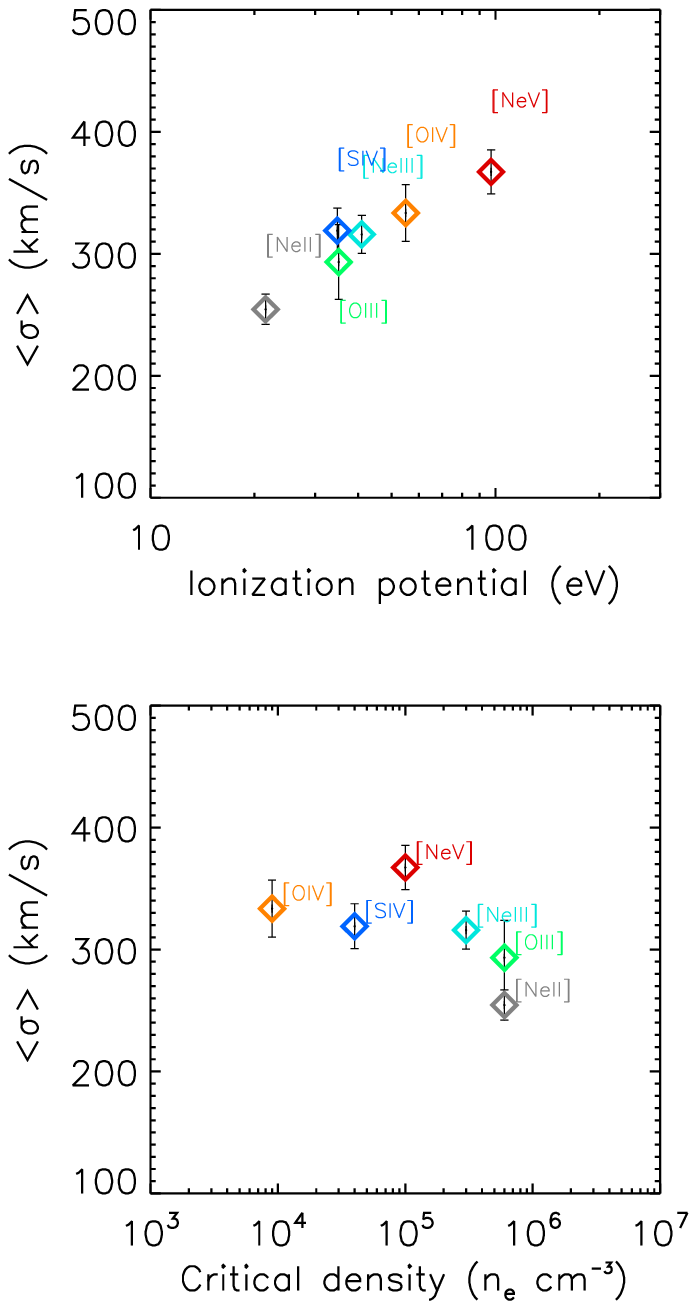}
\caption{\label{fig:ip_density}   \footnotesize{ Average NLR gas velocity dispersion (corrected for instrumental broadening) as a function of ionization potential 
(upper panel) and critical density in units of electrons $n_e$ per cubic centimeter (lower panel) for 16 sources with detections in all of the 
lines used for the construction of this figure.}}
\end{figure}

\begin{figure}[h!]
\centering
\includegraphics[width=8cm]{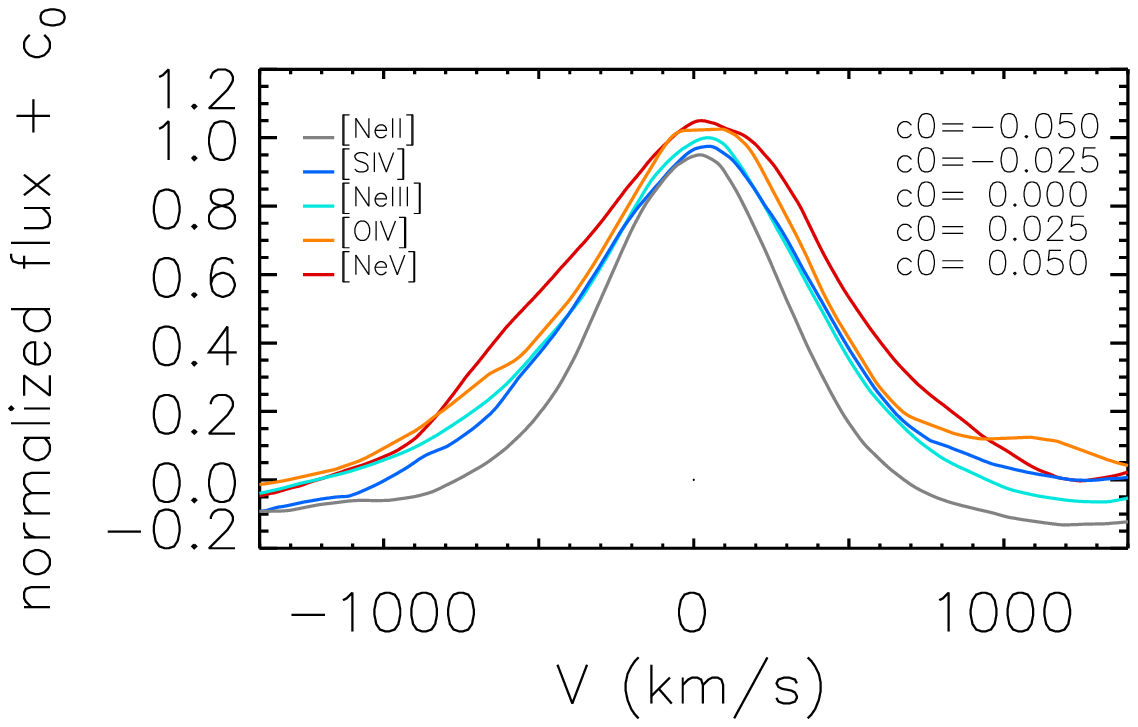}
\caption{\label{fig:stacked_sigma}   \footnotesize{ Stacked MIR line profiles of the 16 sources used in Figure~\ref{fig:ip_density}}.}
\end{figure}

Our query for sources with resolved \siv , \neiii , \oiv, or \nev\  lines resulted in 81 AGN with kinematic information of the NLR gas in the MIR
(Figures~\ref{fig:profiles1} and~\ref{fig:profiles2}). The different fine-structure lines that were used as primary tracers of clouds photoionized by the AGN 
often led to different $\sigma$ measurements (Figures~\ref{fig:profiles1} and~\ref{fig:profiles2}).  Differences in the ionization potential of the various ionic 
species can play a major role in determining the observed line profiles, alongside with differences in the critical density for collisional de-excitation 
of the various transitions, and with light extinction by dust particles. We find the \nev\ line to be often broader than lines from ions of lower ionization potential 
(e.g., Mrk590; Figures~\ref{fig:profiles1}  and~\ref{fig:profiles2}; Tables~\ref{tab:type1_ir} and~\ref{tab:type2_ir}). A significant increase in the average velocity 
dispersion with increasing ionization potential is shown in Figures~\ref{fig:ip_density} and~\ref{fig:stacked_sigma}, as found for 16 sources with resolved 
profiles in all \neii, \oiii , \siv, \neiii, \oiv, and \nev\ lines. This result provides evidence for a stratification of the NLR clouds, i.e., for ions of high ionization 
being located nearer to the BH than ions of low ionization potential (\citealt{filippenko84}; \citealt{oliva94}; \citealt{ho09}).

Radially dependent line widths will also be produced if the NLR has a gas density gradient increasing with proximity to the BH (\citealt{filippenko84};
\citealt{ferguson97}), which is not connected to a mean ionization level gradient. The line widths will be affected as the line fluxes 
from transitions of low critical density for collisional de-excitation, $n$$_{\rm c}$, will be preferentially suppressed at dense environments. We observe 
no trend of $\sigma$ with $n$$_{\rm c}$ (Figure~\ref{fig:ip_density}), suggesting that the typical NLR gas densities are below $n$$_{\rm c}$ for the MIR 
transitions that we examined. The lack of dependence of the average $\sigma$ on $n$$_{\rm c}$ demonstrates the power of MIR lines 
in probing the NLR kinematics. One counter example to this statistical finding could be MCG-03-34-064. Because all of its neon lines have higher 
velocity dispersions than \oiv\ and \siv , its NLR density could be of order 10$^{4}$ hydrogen atoms per cubic centimeter.

Dust either mixed or outside the NLR gas clouds can lead to different extinction of lines at different $\lambda$ (\citealt{groves04}; \citealt{winter10}). 
Because MIR lines are less affected by obscuration than optical lines, they could be probing clouds that are nearer to the BH, where obscuration is 
often higher. In this case, their widths could be broader than those of optical lines, in particular in type 2 AGN. Mrk273, NGC2623, and IRAS15001+1433  
do have larger \oiv\ than \oiii\ widths and low \oiii /\oiv\ flux ratios (see Figure~\ref{fig:line_luminosities}), but their \oiii\ and \siv\ or \neiii\ widths are in good 
agreement with each other. The comparable ionization potentials of  \ion{O}{3}, \ion{S}{4}, and \ion{Ne}{3}, i.e., 35.12, 34.79, and 40.96 eV respectively, 
ascribe again any profile differences to ionization effects. We conclude that obscuration is not a common driver of differences in the profiles of optical and 
IR lines. The nuclear obscuration could be high enough to hide blue wings, associated with outflows moving away from the observer, even in IR wavelengths.

\begin{figure*}
\centering
\includegraphics[width=17cm]{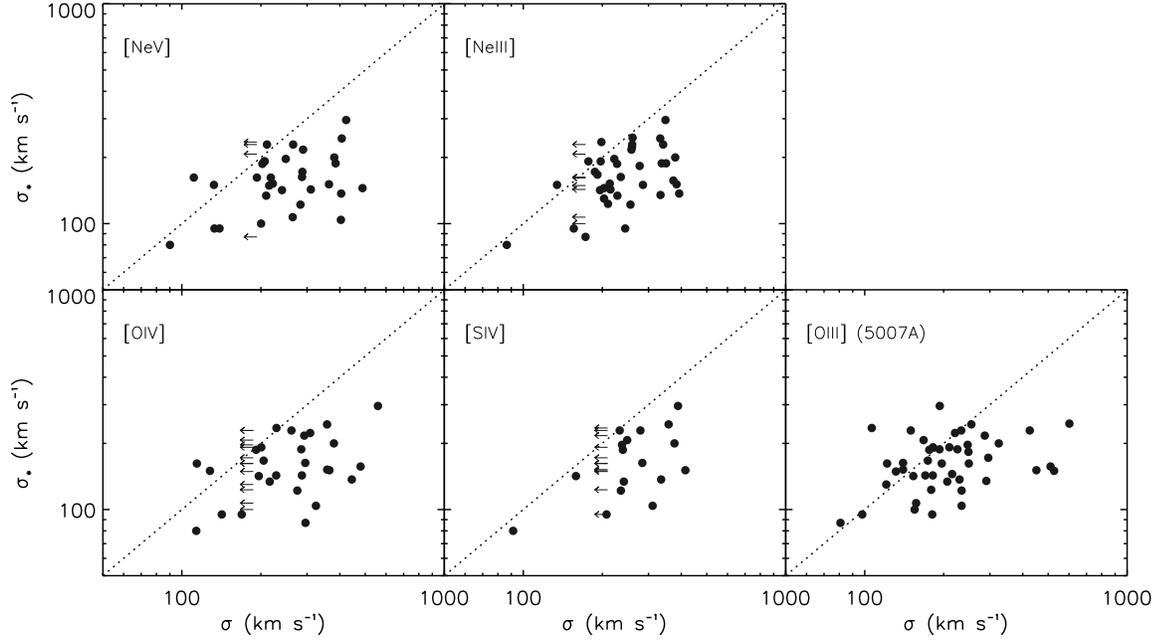}
\caption{\label{fig:nlr_stars}  \footnotesize{ Comparison of the stellar velocity dispersion with the NLR gas velocity dispersion, as measured from MIR lines and from the optical \oiii\ 5007\ang\ line.}}
\end{figure*}

A simple comparison of the NLR gas velocity dispersion to that of the stars in the host galaxy shows that  $\sigma$ is not identical to $\sigma_*$  
(Figure~\ref{fig:nlr_stars}). The gas velocity dispersion systematically exceeds that of the stars, with a large scatter between the two quantities. The 
average velocity dispersion excess is in the range 50-100 \kms\ for all lines. The excess could be intrinsically higher, given that the optical observations 
had a slit width that was often an order of magnitude narrower than the IRS slit width. The origin of this excess could be related to either gas clouds that 
are accelerated by AGN-feedback mechanisms (\citealt{ho09}), or to virialized gas clouds with a more compact spatial distribution than that of the stars,
when probed by emission lines from ions of ionization potential $\gtrsim35$eV. High spatial resolution MIR spectroscopy 
of 8 local AGN indicated that the \siv\ emission is unresolved at a scale of 100 pc (\citealt{hoenig08}), while NIR integral-field-unit data of Circinus showed 
that most of the [\ion{Si}{6}] and [\ion{Ca}{8}] emission is unresolved at 4 pc (\citealt{muller06}). This velocity dispersion excess is seen even for the \oiii\ 
5007\ang\ line, which is resolved for all systems in our sample. \cite{greene05} had also reported such an excess in a large flux-limited sample of AGN 
observed with SDSS.

Another trend known from optical NLR studies is that the widths of the lines increase with their own luminosities  \citep{phillips83,whittle85,whittle92b}
We reproduce this result for the MIR fine-structure lines that we examined in Figure~\ref{fig:lum_vs_width}. The velocity dispersion 
correlates roughly as $L$$^{0.15}$ based on, e.g., the \nev\ measurements. A higher dependence of $\sigma$ on $L$ is plausible when taking into account 
the upper limits of $\sigma$. An exact determination of this slope will require the use of a flux-limited sample of AGN in the local universe.

\subsection{Line Profiles: Shifts and Asymmetries}
\label{sec:results_asymmetries}

Changes in the profile moments of different lines can also be seen when the emission originates from gas out of dynamical equilibrium.  
Several of the 81 sources in our sample show signatures of outflowing (or inflowing) gas motions, such as an offset from systemic velocity. 
This offset often depends on the ionization potential of each ionic species. With increasing ionization potential, the lines trace clouds that are nearer 
to the BH, and therefore more susceptible to outflows. We identified 6 sources in which the offset between the peak of the lowest ionization line 
available and the highest ionization line available exceeds Nyquist sampling of the resolution element, $\gtrsim$250\kms . Namely, these 
are the type 1 AGN 3C273, IRAS13342+3932, and the type 2 AGN IRAS05189-2524, IRAS15001+1433, IRAS23060+0505, and Mrk609 (see also 
\citealt{spoon09a};  \citealt{spoon09b}). Line shifts below Nyquist sampling of the instrumental resolution, yet above the wavelength calibration 
uncertainty (50 \kms ), could also be theoretically observed. They could be considered real if they systematically increased as a function 
of the ionization potential of the ionic species. Only one source\footnote{Other sources, such as NGC2623, NGC7469, 
Arp148, and PG1211+143, could also have outflows of ionized gas, which need to be proven with spectroscopy of higher resolution than that of
IRS. The blue wings of \nev\ and \oiv\ in NGC6240 are also redshifted with respect to \neii\ and \neiii , but their overall line shifts are hard to 
determine as these lines are blended with [\ion{Cl}{2}] and \feii\ (see also \citealt{armus06}). }
satisfies this criterion, Mrk1457. Its \nev\ and \neii\ lines are offset by 100 \kms . 

Asymmetric wings  are found in $\sim$1/5th of the sources with resolved
 profiles. We consider such wings to be reliable only when they are detected in two or more lines. Similar \oiii\ 5007\ang\ and 
\neiii\ or \nev\ wings are observed, for example, in PG1351+640, PG1411+442, and  PG1440+356. Likewise, an agreement of the \oiii\ and 
 \oiv\ 25.89\um\ line profiles is found for 3C273,  IRAS13451+1232, IRAS15001+1433, Mrk1014, NGC4235,  
 and NGC7674. Specifically, IRAS13451+1232 is a merging system with two nuclei separated by $\sim$5 kpc (e.g., \citealt{axon00};
 \citealt{dasyra06a}).  The velocity of the secondary \oiv\ peak , at $\sim$-1000\kms, is comparable to the velocities of the outflowing \oi\ and \oiii\ 
gas components that are seen in optical spectroscopy and that are associated with the nucleus responsible for the radio-jet emission (\citealt{holt03}). 


\section{Discussion }
\label{sec:discussion}

\subsection{A Three Dimensional Plane connecting \mbh , and the $L$ and $\sigma$ of the NLR?}
\label{sec:discussion_mbh_calibration}

By performing a similar profile analysis of the \nev\ and \oiv\ lines, restricted only to AGN with direct \mbh\ measurements from reverberation 
experiments (\citealt{peterson04}), we previously demonstrated that the NLR velocity dispersion correlates with the mass of its central BH 
(\citealt{dasyra08}). We now further populate this relation using data from the full \spi\ archive, complementing them with ISO SWS data to cover the 
parameter space $\sigma$$\lesssim$ 200\kms . The \mbh\ values for this expanded sample include single-epoch optical spectroscopic estimates 
for type 1 AGN, as well as direct \mbh\ measurements for a few type 2 AGN (see Section~\ref{sec:sample}; Tables~\ref{tab:type1_ir} and~\ref{tab:type2_ir}).

\begin{figure*}
\centering
\includegraphics[width=16cm]{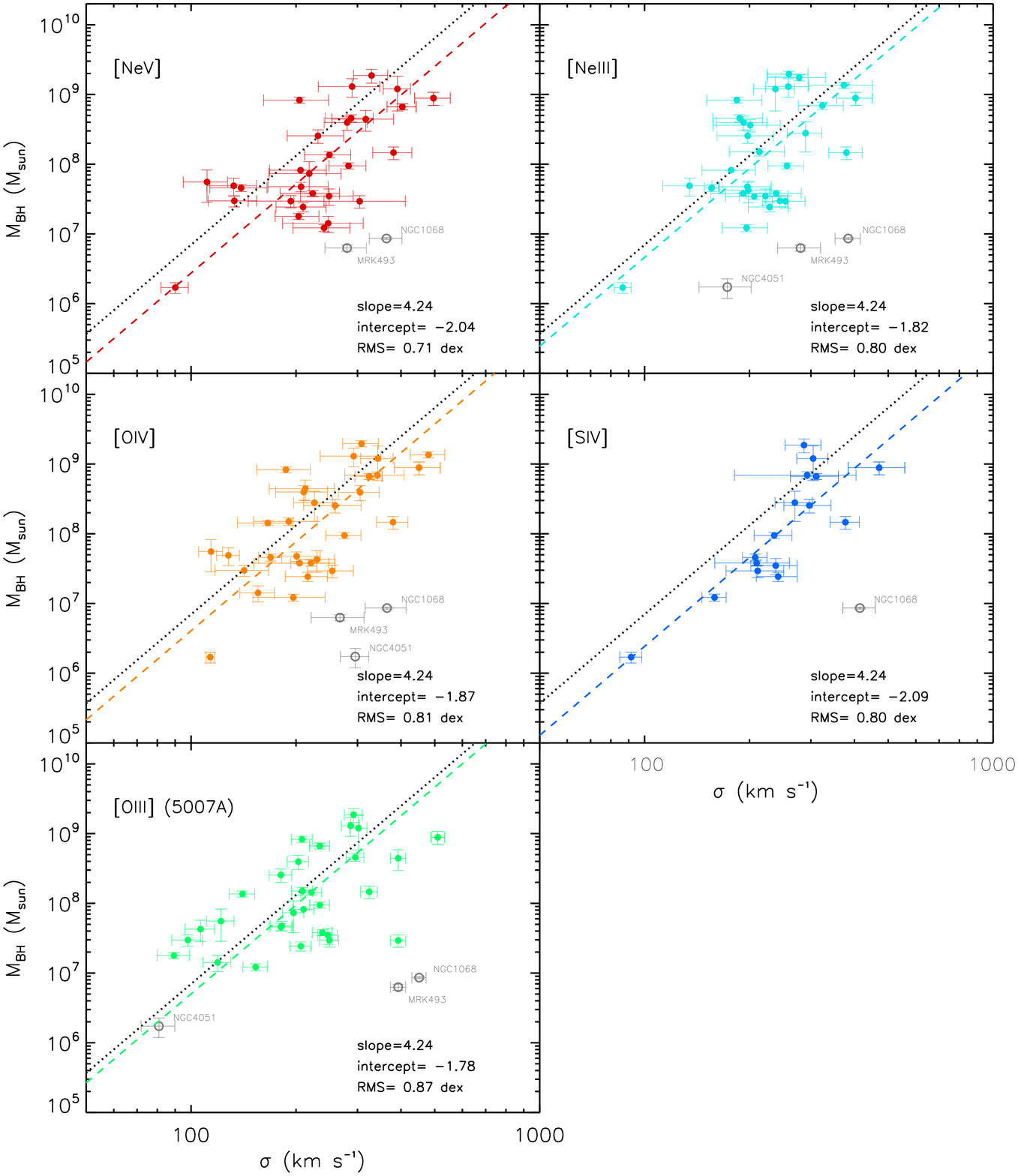}
\caption{\label{fig:msigma}  \footnotesize{  \mbh\ vs NLR velocity dispersion. Open circles correspond to sources included in the rms computation, but excluded from the fit due
to the lack of corresponding \spi\ data for the left-hand side of the relation. The best-fit solution to all other data points is presented with a dashed line. The dotted 
line corresponds to the \cite{gueltekin09} relation, thought appropriate for the velocity dispersion of the stars in the bulge. }}
\end{figure*}

\begin{figure*}
\centering
\includegraphics[width=16cm]{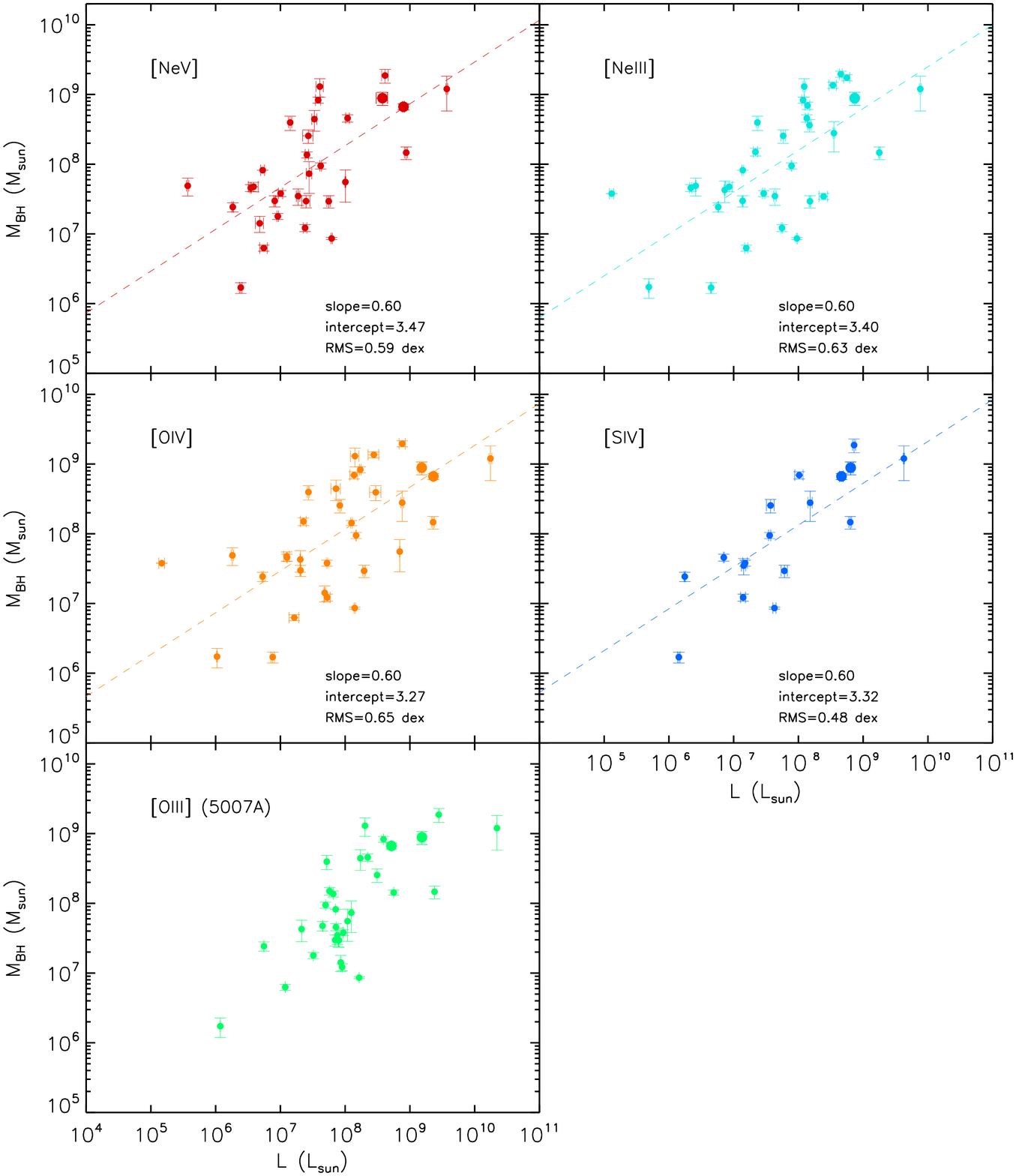}
\caption{\label{fig:mlum}  \footnotesize{ \mbh\ vs narrow line luminosity relation. }}
\end{figure*}

The fit to these data (Figure~\ref{fig:msigma}) was performed with the IDL routine MPFITFUN.  Given the sparsity of data at the low $\sigma$ end, we 
opted for a linear fit of fixed slope in logarithic space. The slope was set to 4.24, which is identical to that of the stellar relation (\citealt{ferrarese00};
\citealt{gebhardt00}) as recently revisited by \cite{gueltekin09}.
While the rms scatter was computed using all available datapoints, the best-fit solution was computed using only \spi\ observations of sources 
with \mbh $>$10$^7$\msun\ or ISO observations, as in \cite{dasyra08}. The reason why we excluded \spi\ datasets for \mbh $<$10$^7$\msun\ 
is that they would introduce a bias toward high intercept values: the resolution of IRS is insufficient to resolve lines on the left-hand side of the 
\msigma\ relation 
below this threshold. This excludes the type 2 AGN NGC1068, and the narrow-line Seyfert 1s (NLS1s) Mrk493, and NGC4051 that have high $\sigma$
values for their BH masses. The existence of such outliers nonetheless suggests that $\sigma$ could fail as a proxy of \mbh , as it is also known for the
BLR gas (e.g., \citealt{vestergaard06}).  On the other hand, such sources are not outliers in the relation connecting \mbh\ to the luminosity 
of the NLR lines (Figure~\ref{fig:mlum}). 

\begin{figure*}
\centering
\includegraphics[width=17cm]{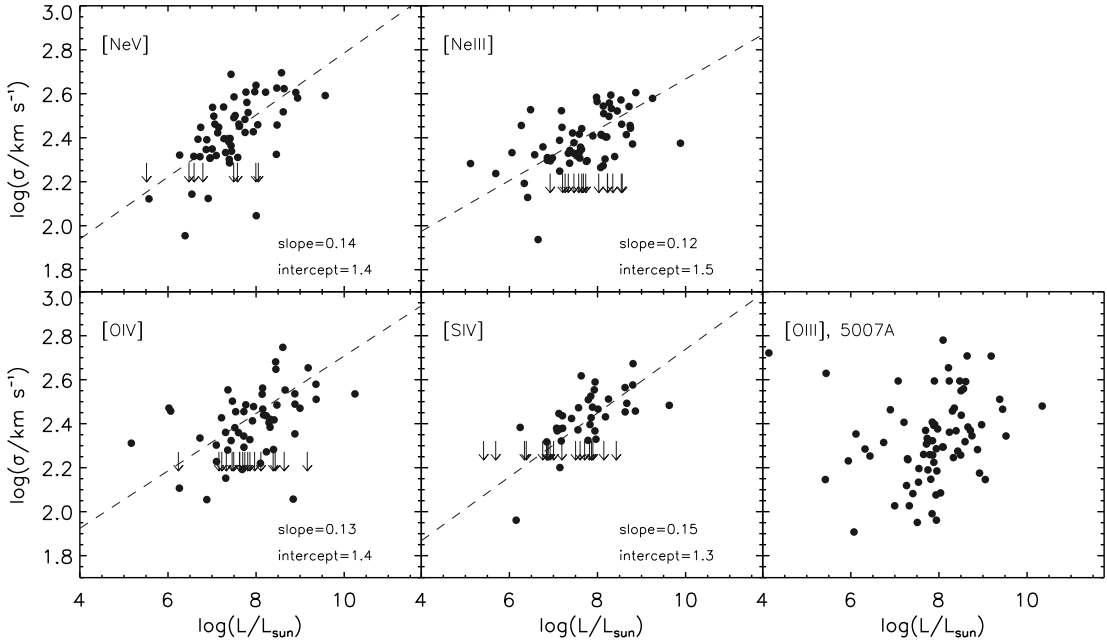}
\caption{\label{fig:lum_vs_width}  \footnotesize{ Velocity dispersion increase with the luminosity of the line from which it was measured. The error-weighed best-fit solution for 
each MIR line is presented with a dashed line.}}
\end{figure*}

\begin{figure*}
\centering
\includegraphics[width=18cm]{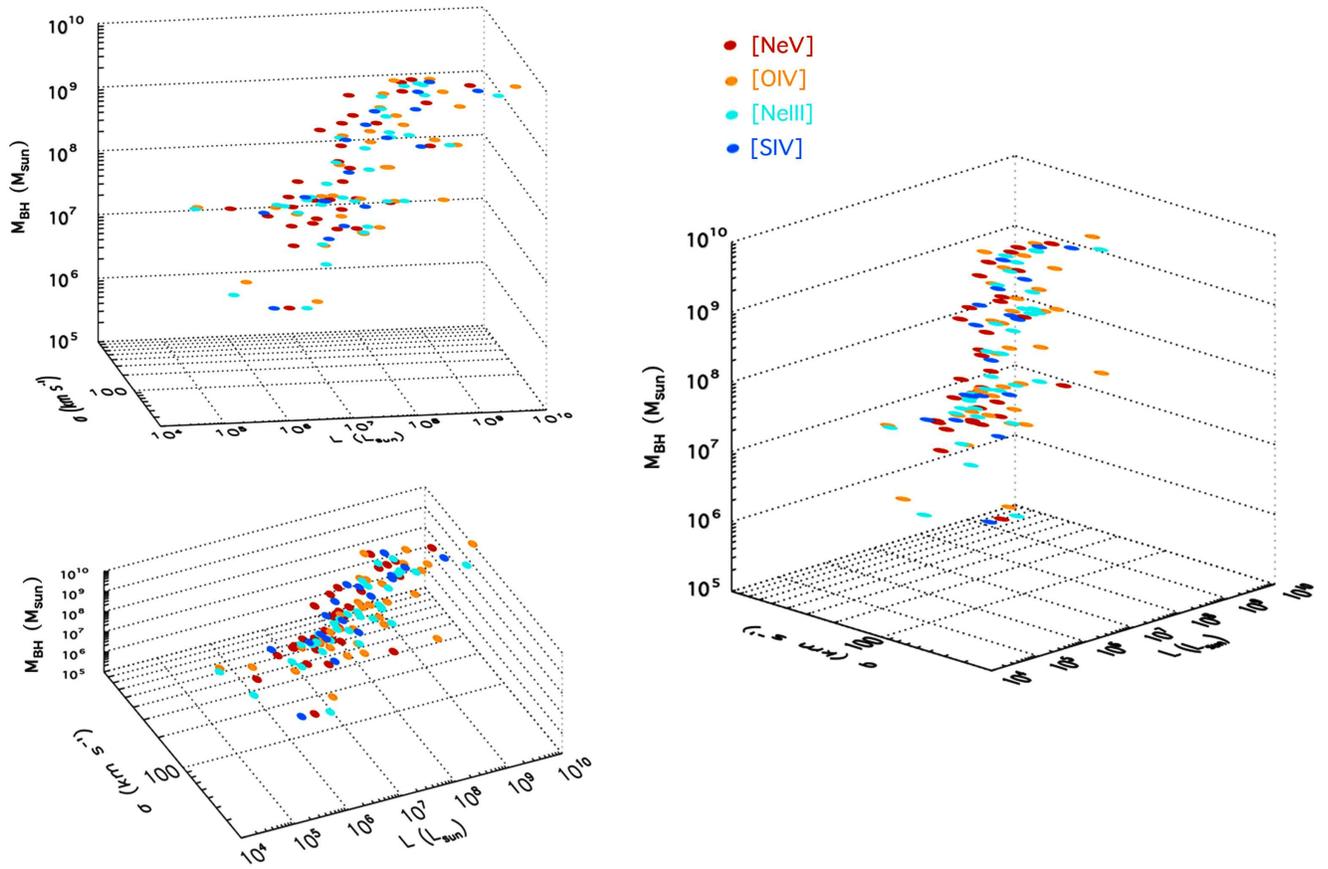}
\caption{\label{fig:plane}  \footnotesize{ Different views of the plane suggested to connect \mbh , and the velocity dispersion and luminosity of AGN narrow lines in the MIR.}}
\end{figure*}

\begin{figure*}
\centering
\includegraphics[width=16cm]{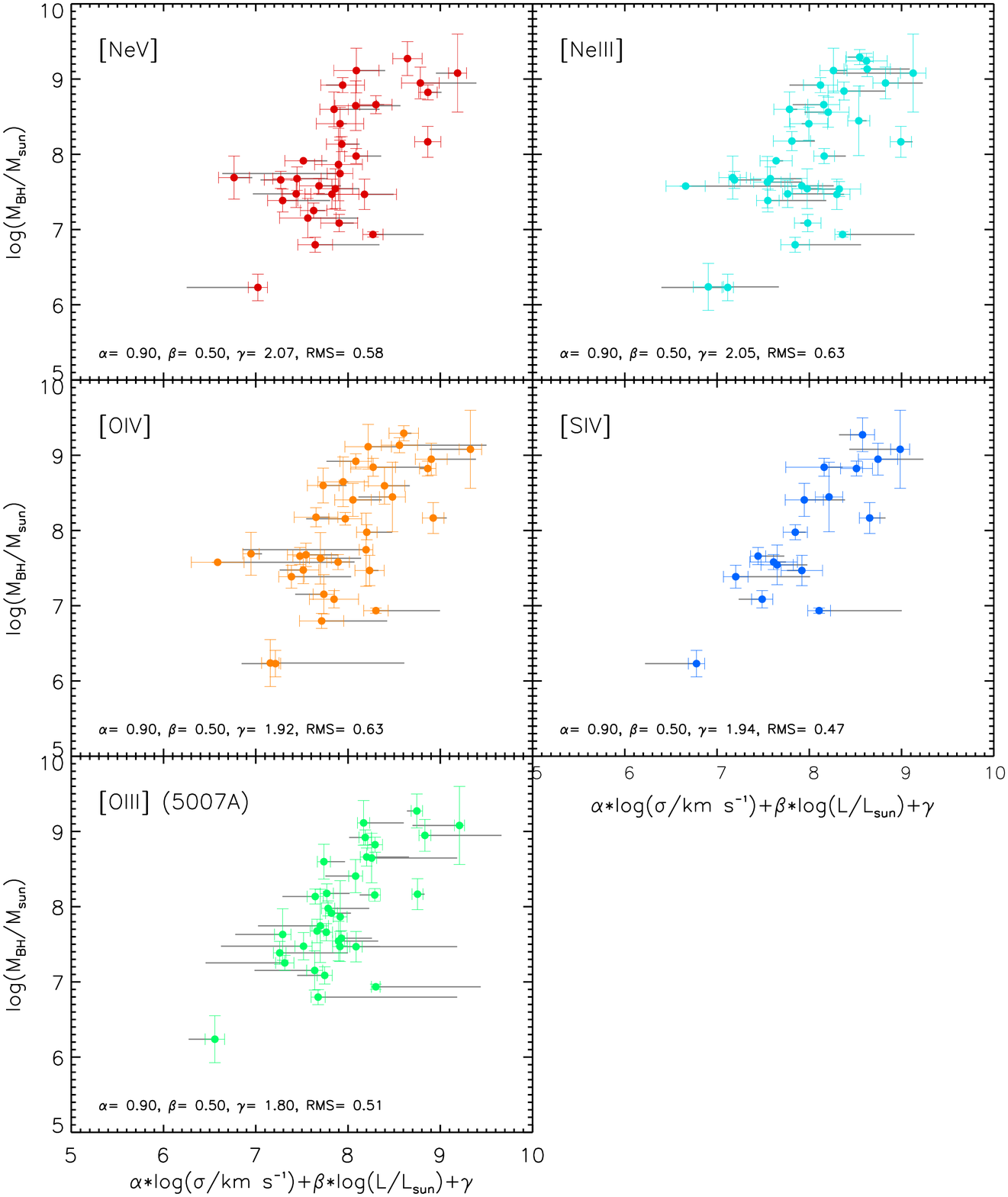}
\caption{\label{fig:mbh_pc}  \footnotesize{ Linear combination, i.e., plane equation, of log($\sigma$) and log(L) that describes log(\mbh). The horizontal solid lines indicate where 
each source would have been located if its \mbh\ was estimated from the \msigma\ relation (Figure~\ref{fig:msigma}) instead.}}
\end{figure*}

The multiple relations between \mbh\ and $\sigma$, $\sigma$ and $L$, and \mbh\ and $L$ could be suggestive of a plane connecting these three parameters
(Figure~\ref{fig:plane}), which we fit using the equation
\begin{equation}
\label{eq:plane}
\footnotesize{ 
log(\mbh) = \alpha log(\sigma)  + \beta log(L)  +  \gamma ,
}
\end{equation}
where $\alpha$,$\beta$, and $\gamma$ are constants. We find that the best-fit-solution coefficients correspond to $\alpha$=0.9 and $\beta$=0.5, when
averaged over all MIR lines. Given the $L$ and $\sigma$$^{\sim7}$ proportionality shown in Figure~\ref{fig:lum_vs_width}, this result roughly reproduces the 
$\mbh \sim \sigma^{4}$ relation. In Figure~\ref{fig:mbh_pc}, we present the plane equation for each line, when fixing the log($\sigma $) and log($L$) slopes 
to their average values for simplicity\footnote{
We have adapted this approach throughout all fitting procedures in this work (see also Figures~\ref{fig:msigma} and \ref{fig:mlum})  to facilitate comparisons 
of the rms scatter among the various relations, and to avoid biases related to the small galaxy sample that is used for their creation and that varies from 
line to line.}. We find that the use of a plane equation minimizes the scatter of the NLR-based \mbh\ estimates from their actual values. The improvement 
primarily originates from the correction of the \msigma\ relation outliers at its low-\mbh\ and high-$\sigma$ end.

\subsection{The Origin of the Scaling Relations between the NLR Gas Properties and \mbh}
\label{sec:discussion_origin}
The physical interpretation of the suggested plane linking the NLR line luminosity, velocity dispersion and the BH mass, varies depending upon the assumed kinematics 
and distribution of the gas clouds. A plane equation can be meaningful for clouds on accelerated motions powered directly by the AGN, via radiation pressure 
(\citealt{murayama98b}) acting mostly upon dust particles due to their high opacity (\citealt{dopita02}), magnetic fields related to jets (\citealt{whittle92c}), 
and AGN-related winds. The winds can lead to either asymmetric outflows, identified in 7\% of the systems in this sample, or to symmetric outflows that can be related to 
the AGN accretion disk (\citealt{crenshaw03}; \citealt{ho09}). In this scenario, the measured gas velocity would be related to the fraction of the energy generated by the
AGN that is deposited into the NLR gas. The luminosity would be a probe of the AGN accretion rate, since the luminosities of MIR NLR lines are known 
to be correlated with the optical, X-ray, and bolometric AGN luminosities (\citealt{schweitzer06}; \citealt{dasyra08};  \citealt{melendez08}; \citealt{rigby09}; R. Mor et al. in 
preparation). A plane connecting $\sigma$, $L$, and \mbh\ would then suggest that the kinetic energy of the interstellar medium (ISM), as measured from the MIR lines, 
is directly related to the amount of material that is accreted onto the BH for a given \mbh\ value. An analogous idea was introduced by \cite{merloni03}, who found a 
correlation between the mass of an accreting BH with its X-ray and radio luminosities.

Instead of responding to AGN feedback mechanisms, the gas clouds could be in virial motions that are dictated by the enclosed mass  at the NLR radius $R$. 
For a geometric factor $f$ converting this total mass to the BH mass, the virial equation can be written as 
\begin{equation}
\label{eq:mbh_virial}
\footnotesize{ 
log(\mbh)= 2 log(\sigma) + log(R) +log(f) -  log(G),
}
\end{equation}
where $G$ is the gravitational constant. In this model,  the NLR radius depends on the AGN luminosity alongside with the stellar mass and distribution in the 
AGN host galaxy. Equation~(\ref{eq:mbh_virial}) will take the form of Equation~(\ref{eq:plane}) if $R$ scales with the AGN luminosity as a power law of $L$, as found
for the \oiii\ 5007\ang\ line (\citealt{bennert02}; \citealt{schmitt03}). The hardness of the ionizing radiation together with the ionization fraction, the covering 
factor, and the density distribution of the clouds can also determine how far the AGN radiation reaches, illuminates, and photoionizes gas clouds. For example, the
observed increase of $\sigma$ with $L$ can be driven either by a tendency for more massive BHs to reside in larger galaxies (\citealt{ho09}), or by the ionization 
state of the NLR. Lines from ions of high ionization potential are thought to be tracing matter-bound clouds that are partially ionized (\citealt{murayama98a}; 
\citealt{wilson97}). In the case of partially ionized clouds, an increase of the AGN luminosity can lead to an overall expansion of the NLR, while most of the
ionization can still occur for clouds at small radii. The exact relation between $\sigma$, $L$, and \mbh\ will depend on the gas density distribution, which is 
encapsulated in the scaling factor $f$. The coefficients of the best-fit plane solution, $\alpha$=0.9 and $\beta$=0.5, suggest that $f$ is dropping with $\sigma$, 
corresponding to a lower scaling factor for denser gas within a fixed radius. If however, the scaling factor $f$ was constant, $\alpha$ would be equal to 2, 
and the best-fit plane solution would correspond to $\beta$=0.4. Such a solution could also be plausible. It would increase the rms scatter by a small amount, 
i.e., by only 0.01$-$0.02 dex for the MIR lines and by 0.07 dex for \oiii.

If the cloud kinematics cannot be approximated by either feedback-driven motions or by virial motions, Equation~\ref{eq:plane} might not provide a good means of estimating 
\mbh . For cloud kinematics that are described by a combination of virial and accelerated motions, e.g., in a virialized NLR where radiation pressure is significant, \mbh\ 
is proportional to $\sigma^2 R G^{-1} + a' L$, where $a'$ is a constant that depends on the column density and level of ionization of the clouds (\citealt{marconi08}; 
but see also \citealt{netzer10}). Further investigation of the optimal description of \mbh\ from narrow line properties will be tested on large, flux-limited samples of AGN 
with multiwavelength datasets. This will be the focus of a forthcoming paper using SDSS data.

\subsection{Black Hole Mass Estimates in Obscured AGN}
Given that we see no significant difference in the NLR properties of type 1 and type 2 AGN at a given $L$ or $\sigma_*$ (see also \citealt{pereira10}), we applied the 
relations presented earlier in this work to type 2 AGN for which such an analysis is presently possible from \spi\ data. Only four of these sources have other, direct 
\mbh\ measurements (Table~\ref{tab:type2_sum}). The comparison between the \mbh\ estimates from the stellar \msigma$_*$ relation, the \msigma\ relation for the 
NLR gas, and  the best-fit 
linear combination of log($\sigma $) and log($L$) are presented in Table~\ref{tab:mbh2}.  For either computation based on the NLR gas, the result is averaged over all 
available MIR lines. We find that the median ratio of \mbh\ as estimated from the plane equation over \mbh\ as estimated from the stellar velocity dispersion is 1.8. On 
the other hand, the median ratio of \mbh\ as estimated from the NLR gas velocity dispersion over \mbh\ as estimated from the stellar velocity dispersion is much 
higher, 6.5. The difference seen when folding a luminosity dependence on the computation of \mbh\ is largest for the IR-bright galaxies IRAS05189-2524, 
IRAS15001+1433, and Mrk609, which have gas motions that are predominantly out of dynamical equilibrium.


\section{Conclusions}
\label{sec:conclusions}

We queried the full \spi\ archive for high-resolution IRS spectra of type 1 AGN with BLR-based \mbh\ estimates from optical spectroscopy, 
and type 2 AGN. We analyzed the spectra of 298 objects, which we combined with ISO spectra of 6 more 
AGN, aiming to study the fundamental properties of the NLR in the IR, and to calibrate its fine-structure-line widths and luminosities to the 
mass of the central black hole. We found the following. 
\begin{itemize}
\item
All of the \nev , \oiv , \neiii , and \siv\ line luminosities are well correlated with each other (with an rms scatter of $\sim$0.2 dex). 
Given the high ionization potential of \ion{Ne}{4} to \ion{Ne}{5}, their emission is attributed to (further) photoionization of ions by 
the AGN. Thus, all of these lines can be used to study the NLR kinematics.
\item
The bulk of the silicates responsible for the 9.7 \um\ absorption feature is likely to be located in a region smaller than the NLR. 
Otherwise, the luminosity of \siv\ at 10.51 \um\ would not correlate equally well with that of all other IR narrow lines without an 
extinction correction in obscured AGN.
\item
Of the 304 sources in our sample, 81 had at least one \nev , \oiv, \neiii , or \siv\  narrow line that was resolved. We find that, on 
average, the line widths increase with increasing ionization potential of the species that emit them in the range 22$-$97eV. Different 
fine-structure lines probe different (locations within the) clouds, with high-ionization-potential ions being preferentially found 
nearer to the BH. 
\item
No trend was observed between the line width and critical density for these transitions, indicating that the average density of 
the NLR gas is typically below 10$^4$ hydrogen atoms per cubic centimeter.
\item
The velocity dispersions of all MIR lines that we examined, as well as that of the \oiii\ 5007\ang\ line, are systematically above the
velocity dispersion of the stars in the AGN host galaxies.  Moreover, the line widths increase with their own luminosities, as known from 
optical wavelengths. 
\item
The mass of the central BH correlates with the velocity dispersion and with the luminosity of the NLR lines. We speculate that
these results could be suggestive of a three-dimensional plane, connecting log($\mbh $) to a linear combination of log($\sigma $) and log($L$). 
Such a plane equation could be meaningful for a virial distribution in which the NLR radius has a power-law dependence on the AGN luminosity, 
or for AGN-feedback driven motions.
\item
Given that we find no significant differences in the NLR kinematic properties of type 1 and type 2 AGN, we used the NLR gas properties as seen in 
the MIR to estimate the BH masses residing in 35 local obscured AGN.
\end{itemize}

\acknowledgments
 K. D. acknowledges support by the European Community through a Marie Curie Fellowship (PIEF-GA-2009-235038) 
 awarded under the Seventh Framework Programme (FP7/2007-2013). The authors wish to thank T. Boroson and B. Peterson 
 for providing optical spectra to be compared with the MIR ones, and S. Collin for useful discussions. This work made use of the 
 NASA Extragalactic Database, and it was based on archival data obtained with the \spi\  Space Telescope, operated by the Jet 
 Propulsion Laboratory, California Institute of Technology under a contract with NASA.


{\begin{onecolumn}

\clearpage
\oddsidemargin=0in 
\begin{centering}
\begin{deluxetable}{cccccc}
\tablecolumns{6}
\tabletypesize{\tiny}
\tablewidth{0pt}
\tablecaption{\label{tab:type1_sum} Black hole masses \& optical NLR and host galaxy properties of type-1 AGNs.}

\tablehead{ 
\colhead{Galaxy} & \colhead{$z$} & \colhead{\mbh } & \colhead{log(L$_{\rm {[OIII]}}$/(ergs s$^{-1}$))} & \colhead{\soiii (5007\ang)} & \colhead{\sstar} \\
\colhead{(-)} & \colhead{(-)} & \colhead{($10^7$ \msun )}  & \colhead{(-)} & \colhead{(km s$^{-1}$) } & \colhead{(km s$^{-1}$) } \\
\colhead{(1)} & \colhead{(2)} & \colhead{(3)}  & \colhead{(4)} & \colhead{(5)} & \colhead{(6) } \\
}
\startdata

                      2MASSJ09184900+2117170 &   0.1490 &     3.45 $\pm$     0.35 & 42.09 &  274 & \nodata \\ 
                       2MASSiJ1659397+183436 &   0.1709 &   174.28 $\pm$    17.43 & 42.55 &  249 &  183 $\pm$  10 \\ 
                      2MASXJ10514428+3539304 &   0.1588 & \nodata & 41.67 &  197 & \nodata \\ 
                                     3C249.1 &   0.3110 &   187.10 $\pm$    42.20 & 43.03 &  292 & \nodata \\ 
                                       3C273 &   0.1583 &    88.60 $\pm$    18.70 & 42.77 &  510 & \nodata \\
                                      Ark120 &   0.0328 &    15.00 $\pm$     1.90 & 41.34 &  209 & \nodata \\ 
                                 CGCG121-075 &   0.0323 &     3.81 $\pm$     0.38 & 41.56 &  238 & \nodata \\ 
                                 ESO140-G043 &   0.0142 &     1.79 $\pm$     0.18 & 41.09 &   89 & \nodata \\ 
                                    FAIRALL9 &   0.0465 &    25.50 $\pm$     5.60 & 42.08 &  181 & \nodata \\ 
                                     IC4329A &   0.0160 &     9.47 $\pm$     0.95 & 41.28 &  234 &  122 $\pm$  13 \\
                              IRAS13342+3932 &   0.1793 &    66.61 $\pm$     6.66 & 42.30 &  234 &  104 $\pm$  14 \\ 
                              IRAS13451+1232 &   0.1217 &   135.96 $\pm$    13.60 & 42.22 &  511 &  157 $\pm$  39 \\ 
                              IRAS18216+6418 &   0.2970 &   120.00 $\pm$    62.20 & 43.93 &  302 & \nodata \\ 
                                 MCG-2-58-22 &   0.0469 &    83.13 $\pm$     8.31 & 42.18 &  208 & \nodata \\ 
                                     Mrk1014 &   0.1631 &    14.66 $\pm$     3.01 & 42.96 &  324 &  200 $\pm$  60 \\
                                      Mrk279 &   0.0305 &     3.49 $\pm$     0.92 & 41.47 &  247 &  197 $\pm$  12 \\  
                                      Mrk335 &   0.0258 &     1.42 $\pm$     0.37 & 41.52 &  119 & \nodata \\ 
                                      Mrk493 &   0.0313 &     0.63 $\pm$     0.06 & 40.66 &  393 & \nodata \\ 
                                      Mrk590 &   0.0264 &     4.75 $\pm$     0.74 & 41.24 &  182 &  192 $\pm$  10 \\ 
                                      Mrk704 &   0.0295 &    13.64 $\pm$     1.36 & 41.40 &  140 & \nodata \\ 
                                      Mrk705 &   0.0290 &    39.63 $\pm$     9.18 & 41.30 &  203 & \nodata \\ 
                                     NGC3227 &   0.0039 &     2.43 $\pm$     0.37 & 40.33 &  206 &  134 $\pm$   6 \\
                                     NGC3516 &   0.0088 &     4.27 $\pm$     1.46 & 40.91 &  106 &  235 \\ 
                                     NGC4051 &   0.0023 &     0.17 $\pm$     0.05 & 39.66 &   81 &   87 $\pm$   5 \\ 
                                     NGC4235 &   0.0080 & \nodata & 39.53 &  170 &  143 $\pm$  34 \\  
                                     NGC5548 &   0.0172 &     8.20 $\pm$     0.00 & 41.44 &  210 &  192 $\pm$  15 \\ 
                                     NGC7469 &   0.0165 &     1.22 $\pm$     0.14 & 41.54 &  153 &  142 $\pm$   3 \\
                                  PG0026+129 &   0.1455 &    39.30 $\pm$     9.60 & 42.46 &  191 & \nodata \\ 
                                  PG0804+761 &   0.1010 &    69.30 $\pm$     8.30 & 42.06 &  393 & \nodata \\  
                                  PG1119+120 &   0.0502 &     2.95 $\pm$     0.59 & 41.49 &  249 &  162 $\pm$  28 \\ 
                                  PG1229+204 &   0.0630 &     7.32 $\pm$     3.52 & 41.68 &  197 &  162 $\pm$  32 \\  
                                  PG1351+640 &   0.0882 &    36.30 $\pm$     7.26 & 42.19 &  390 & \nodata \\ 
                                  PG1411+442 &   0.0896 &    44.30 $\pm$    14.60 & 41.82 &  393 & \nodata \\ 
                                  PG1426+015 &   0.0866 &   129.80 $\pm$    38.50 & 41.89 &  287 &  217 $\pm$  15 \\ 
                                  PG1440+356 &   0.0790 &     2.94 $\pm$     0.59 & 41.48 &  393 & \nodata \\ 
                                  PG1613+658 &   0.1295 &    27.90 $\pm$    12.90 & 42.08 &  354 & \nodata \\  
                                  PG2130+099 &   0.0630 &    45.70 $\pm$     5.50 & 41.93 &  296 &  172 $\pm$  46 \\  
                                  PG2349-014 &   0.1742 &   196.00 $\pm$    19.60 & 43.11 &  221 &  223 $\pm$  36 \\  
                                  \hline	 
                                       3C120$^*$   &   0.0330 &     5.55 $\pm$     2.70 & 41.62 &  122 &  162 $\pm$  24 \\ 
                                    Mrk509$^*$  &   0.0344 &    14.30 $\pm$     1.20 & 42.33 &  221 & \nodata \\                              
                                    NGC3783$^*$  &   0.0097 &     2.98 $\pm$     0.54 & 41.43 &   98 &   95 $\pm$  10 \\
                                  NGC4151$^*$  &   0.0033 &     4.57 $\pm$     0.52 & 41.44 &  181 &   95 $\pm$   7 \\
\enddata

\tablecomments{ ($^*$) sources with MIR fine-structure lines resolved only by ISO.\\
Col. (3): When possible, \mbh\ measurements are taken from reverberation mapping experiments (\citealt{peterson04}; \citealt{bentz06}; \citealt{bentz09b}; \citealt{denney10}). 
Otherwise, they are taken from single-epoch spectroscopy (\citealt{vestergaard06}; \citealt{kim08}) or computed from single-epoch spectroscopy (\citealt{marziani03}; 
 \citealt{netzer07}; \citealt{hokim09}; SDSS archival spectra) as in \cite{vestergaard06}.\\
2MASXJ10514428+3539304 has weak, broad \ha\ detection but no \hb\ detection. \\
Col. (4):  Data from \cite{whittle92}, \cite{marziani03}, \cite{hokim09}, SDSS DR7, as well as our own measurements from the spectra presented in \cite{boroson92}, performed as in \cite{netzer07}. The luminosities are not corrected for extinction.\\
Col.(5): \soiii\ values are taken from \cite{whittle92}, \cite{hokim09}, or measured from SDSS spectra.\\
Col. (6): Host galaxy velocity dispersion measurements are taken from \cite{nelson04}, \cite{onken04}, Dasyra et al. (2006a; 2006b; 2007),  \cite{ho07}, and the 
7th data release of SDSS.  When more than one measurements were available, their average value was used. 
IRAS13451+1232 is a merger (\citealt{dasyra06a}). The velocity dispersion presented here corresponds to the combined value for both nuclei.  \\
}
\end{deluxetable}
\end{centering}

\clearpage
\oddsidemargin=0in 
\begin{centering}
\begin{deluxetable}{cccccc}
\tablecolumns{6}
\tabletypesize{\tiny}
\tablewidth{0pt}
\tablecaption{\label{tab:type2_sum} Black hole masses \& optical NLR and host galaxy properties of type-2 AGNs.}

\tablehead{ 
\colhead{Galaxy} & \colhead{$z$} & \colhead{\mbh } & \colhead{log(L$_{\rm {[OIII]}}$/(ergs s$^{-1}$))} & \colhead{\soiii (5007\ang)} & \colhead{\sstar} \\
\colhead{(-)} & \colhead{(-)} & \colhead{($10^7$ \msun )}  & \colhead{(-)} & \colhead{(km s$^{-1}$) } & \colhead{(km s$^{-1}$) } \\ 
\colhead{(1)} & \colhead{(2)} & \colhead{(3)}  & \colhead{(4)} & \colhead{(5)} & \colhead{(6) } 
}
\startdata
                      2MASXJ08035923+2345201 &   0.0297 & \nodata & 40.99 &  121 &  130 $\pm$   5 \\  
                      2MASXJ08244333+2959238 &   0.0254 & \nodata & 41.13 &  157 &  107 $\pm$   5 \\ 
                      2MASXJ10181928+3722419 &   0.0497 & \nodata & 41.33 &  155 &  100 $\pm$   6 \\ 
                      2MASXJ12384342+0927362 &   0.0829 & \nodata & 41.97 &  233 &  229 $\pm$  10 \\ 
                      2MASXJ16164729+3716209 &   0.1518 & \nodata & 42.51 &  150 &  229 $\pm$   9 \\  
                                 CGCG218-007 &   0.0273 & \nodata & 40.85 &  131 &  149 $\pm$   5 \\ 
                                 ESO103-G035 &   0.0133 & \nodata & \nodata & \nodata & \nodata \\ 
                              IRAS05189-2524 &   0.0430 & \nodata & 41.82 &  230 &  137 $\pm$  16 \\ 
                              IRAS15001+1433 &   0.1623 & \nodata & 41.52 &  193 &  296 $\pm$  19 \\ 
                              IRAS18325-5926 &   0.0202 & \nodata & \nodata & \nodata & \nodata \\ 
                              IRAS23060+0505 &   0.1730 & \nodata & 42.24 &  243 & \nodata \\ 
                               MCG-03-34-064 &   0.0165 & \nodata & \nodata & \nodata & \nodata \\ 
                                     Mrk1066 &   0.0120 & \nodata & 40.88 &  172 & \nodata \\  
                                     Mrk1457 &   0.0486 & \nodata & 41.38 &  182 &  143 $\pm$   6 \\  
                                      Mrk273 &   0.0378 & \nodata & 40.79 &  255 &  244 $\pm$  18 \\ 
                                        Mrk3 &   0.0135 & \nodata & 42.14 &  362 & \nodata \\ 
                                     Mrk463E &   0.0510 & \nodata & 42.64 &  140 &  163 $\pm$   8 \\  
                                      Mrk609 &   0.0345 & \nodata & 41.32 &  215 &  145 $\pm$   4 \\ 
                                      Mrk622 &   0.0232 & \nodata & 40.48 &  291 &  135 $\pm$   5 \\ 
                                     NGC1068 &   0.0038 &     0.86 $\pm$     0.03 & 41.80 &  451 &  151 $\pm$   7 \\ 
                                     NGC1275 &   0.0174 & \nodata & 41.68 & 601 &  246 \\ 
                                     NGC2622 &   0.0286 & \nodata & 41.43 &  255 & \nodata \\ 
                                     NGC2623 &   0.0185 & \nodata & 39.01 &  140 &  152 $\pm$   9 \\ 
                                     NGC2639$^{*}$ &   0.0111 & \nodata & 39.70 & 226 &  188  \\
                                     NGC3079$^{*}$ &   0.0037 & \nodata & 37.72 & 527 &  150 \\
                                     NGC4258 &   0.0018 &     3.78 $\pm$     0.01 & 40.87 & 174 &  167 \\ 
                                     NGC4507 &   0.0118 & \nodata & 41.53 &   91 & \nodata \\ 
                                     NGC5256 &   0.0279 & \nodata & 41.91 &  176 &  187 $\pm$   8 \\ 
                                     NGC5506 &   0.0062 & \nodata & 40.58 &  106 & \nodata \\  
                                     NGC5728 &   0.0094 & \nodata & 41.12 &  136 & \nodata \\ 
                                     NGC5929 &   0.0083 & \nodata & 40.02 &  179 &  123 $\pm$   3 \\  
                                     NGC6240 &   0.0245 & \nodata & 39.02 &  426 &  229 $\pm$  43 \\ 
                                     NGC7172 &   0.0087 & \nodata & \nodata & \nodata & \nodata \\ 
                                     NGC7674 &   0.0290 & \nodata & 42.01 &  189 & \nodata \\ 
                                 SBS1133+572 &   0.0516 & \nodata & 41.47 &  168 &  207 $\pm$   7 \\  
                                    UGC02608 &   0.0233 & \nodata & \nodata & \nodata & \nodata \\ 
                                     UGC5101 &   0.0394 & \nodata & 39.91 &  193 &  188 $\pm$   6 \\ 
                                     \hline
                        Centaurus A$^{**}$ &   0.0018 &     4.90 $\pm$     1.40 & \nodata & \nodata &  150 $\pm$   7 \\
                        			    Circinus$^{**}$ &   0.0014 &     0.17 $\pm$     0.03 & 38.55 & \nodata &   80  \\
\enddata
\tablecomments{ ($^{*}$) sources with shallow optical spectra, or spectra obtained under non photometric conditions.\\
($^{**}$) sources with MIR fine-structure lines resolved only by ISO.\\
Col. (3): \mbh\ measurements taken from \cite{lodato03}, \cite{greenhill03}, \cite{herrnstein05}, and \cite{neumayer07}. When more
than one measurements were available, we used the value with the lower uncertainty.\\
Col.(4): The luminosities are computed from \cite{whittle92}, \cite{oliva94}, Veilleux et al. (1995; 1999), \cite{ho97a},  \cite{marziani03} or measured 
from SDSS spectra. They are not corrected for extinction.\\
Col.(5): \soiii\ values taken from \cite{heckman83}, \cite{whittle92},  \cite{veilleux99}, or measured from SDSS spectra.\\
Col. (6): Host galaxy velocity dispersions are compiled from \cite{pahre99}, \cite{tecza00}, \cite{tacconi02}, \cite{dasyra06b}, 
\cite{falcon06}, \cite{hinz06}, \cite{muller06}, \cite{ho07}, \cite{cappellari09}, \cite{gueltekin09}, and from the 7th data release of SDSS. \\ 
}
\end{deluxetable}
\end{centering}

\clearpage
\begin{centering}
\begin{deluxetable}{ccccccccccc}
\rotate
\tablecolumns{11}
\tabletypesize{\tiny}
\tablewidth{0pt}
\tablecaption{\label{tab:type1_ir} Fluxes and Velocity Dispersions of Ionized Gas Narrow Lines in type 1 AGNs.}
\tablehead{ 

\colhead{Galaxy} & \colhead{f$_{\rm [Ne~\sc{II}]} $} & \colhead{f$_{\rm[S~\sc{IV}]} $} & \colhead{f$_{\rm [Ne~\sc{III}]} $} & \colhead{f$_{\rm [O~\sc{IV}]} $} & \colhead{f$_{\rm [Ne~\sc{V}]} $}  & \colhead{\sneii} & \colhead{\ssiv} & \colhead{\sneiii} & \colhead{\soiv} & \colhead{\snev}  \\
\colhead{(-)}  & \colhead{(10$^{-18}$ W m$^{-2}$)} & \colhead{(10$^{-18}$ W m$^{-2}$)} & \colhead{(10$^{-18}$ W m$^{-2}$)} & \colhead{(10$^{-18}$ W m$^{-2}$)} & \colhead{(10$^{-18}$ W m$^{-2}$)} & \colhead{(km s$^{-1}$) } & \colhead{(km s$^{-1}$) } & \colhead{(km s$^{-1}$) }  & \colhead{(km s$^{-1}$)  }  & \colhead{(km s$^{-1}$)  }  
}

\startdata
          2MASSJ09184900+2117170 &             $<$    3.61 &             $<$   10.86 &    15.65 $\pm$     2.39 &                \nodata  &             $<$   25.07 & \nodata & \nodata &  206 $\pm$   35 & \nodata & \nodata \\ 
           2MASSiJ1659397+183436 &             $<$   10.67 &             $<$   10.16 &    26.79 $\pm$     2.95 &                \nodata  &             $<$   10.15 & \nodata & \nodata &  277 $\pm$   52 & \nodata & \nodata \\ 
          2MASXJ10514428+3539304 &             $<$    6.21 &     9.97 $\pm$     1.98 &    19.45 $\pm$     2.11 &                \nodata  &             $<$    5.20 & \nodata &  325 $\pm$   48 & \nodata & \nodata & \nodata \\ 
                         3C249.1 &     4.37 $\pm$     0.40 &     8.85 $\pm$     0.43 &                \nodata  &                \nodata  &     5.05 $\pm$     0.59 &  370 $\pm$   65 &  286 $\pm$   33 & \nodata & \nodata &  329 $\pm$   36 \\ 
                           3C273 &    11.42 $\pm$     1.67 &    36.04 $\pm$     3.33 &    41.58 $\pm$     2.61 &    85.45 $\pm$     6.42 &    21.20 $\pm$     4.21 &  230 $\pm$   39 &  470 $\pm$   87 &  402 $\pm$   47 &  450 $\pm$   67 &  495 $\pm$   57 \\ 
                          Ark120 &    28.97 $\pm$     2.53 &             $<$   13.56 &    35.23 $\pm$     2.71 &    36.84 $\pm$     3.70 &    10.04 $\pm$     1.61 &  202 $\pm$   32 & \nodata &  213 $\pm$   38 &  190 $\pm$   39 & \nodata \\ 
                     CGCG121-075 &    24.88 $\pm$     1.44 &    24.39 $\pm$     2.00 &    47.28 $\pm$     1.96 &    85.25 $\pm$     1.86 &    16.36 $\pm$     1.25 & \nodata &  209 $\pm$   50 &  238 $\pm$   42 &  220 $\pm$   27 &  223 $\pm$   43 \\ 
                     ESO140-G043 &    96.32 $\pm$     2.92 &    87.41 $\pm$     2.95 &   139.5 $\pm$     2.2 &   247.6 $\pm$     4.9 &    79.10 $\pm$     2.19 & \nodata & \nodata & \nodata & \nodata &  203 $\pm$   29 \\ 
                        FAIRALL9 &    21.50 $\pm$     1.85 &    28.59 $\pm$     3.86 &    44.80 $\pm$     3.44 &    63.73 $\pm$     3.68 &    20.79 $\pm$     3.77 & \nodata &  296 $\pm$   45 &  197 $\pm$   36 &  258 $\pm$   48 &  231 $\pm$   42 \\ 
                         IC4329A &   241.5 $\pm$     6.2 &   245.7 $\pm$     9.5 &   535.0 $\pm$     6.8 &  1009 $\pm$    14 &   286.1 $\pm$     8.4 &  227 $\pm$   32 &  235 $\pm$   29 &  256 $\pm$   30 &  275 $\pm$   31 &  282 $\pm$   34 \\ 
                  IRAS13342+3932 &    53.68 $\pm$     1.11 &    19.74 $\pm$     1.27 &             $<$   28.42 &    98.54 $\pm$     1.72 &    33.98 $\pm$     0.98 & \nodata &  310 $\pm$   48 & \nodata &  324 $\pm$   29 &  403 $\pm$   38 \\ 
                  IRAS13451+1232 &    45.66 $\pm$     1.82 &             $<$    5.59 &             $<$   70.31 &    27.87 $\pm$     5.24 &             $<$    7.86 &  311 $\pm$   37 & \nodata & \nodata &  479 $\pm$   54 & \nodata \\ 
                  IRAS18216+6418 &    27.52 $\pm$     2.06 &    58.26 $\pm$     1.23 &   104.5 $\pm$     3.5 &   240.1 $\pm$    12.3 &    50.79 $\pm$     1.69 &  214 $\pm$   34 &  304 $\pm$   31 &  237 $\pm$   31 &  343 $\pm$   36 &  390 $\pm$   34 \\ 
                     MCG-2-58-22 &    74.71 $\pm$     1.86 &             $<$   33.31 &    87.08 $\pm$     3.79 &   126.3 $\pm$     3.6 &    28.26 $\pm$     2.63 & \nodata & \nodata &  183 $\pm$   33 &  187 $\pm$   32 &  204 $\pm$   43 \\ 
                         Mrk1014 &    58.26 $\pm$     1.42 &    33.20 $\pm$     1.67 &    93.20 $\pm$     2.96 &   119.9 $\pm$     5.8 &    46.01 $\pm$     2.39 &  253 $\pm$   37 &  376 $\pm$   36 &  379 $\pm$   41 &  379 $\pm$   39 &  380 $\pm$   48 \\ 
                          Mrk279 &    81.42 $\pm$     2.88 &    26.48 $\pm$     2.04 &    80.35 $\pm$     3.97 &   102.8 $\pm$     3.6 &    34.80 $\pm$     2.19 &  215 $\pm$   34 &  237 $\pm$   36 &  222 $\pm$   35 & \nodata &  248 $\pm$   42 \\ 
                          Mrk335 &    10.54 $\pm$     0.90 &             $<$   25.80 &    22.56 $\pm$     1.75 &   130.0 $\pm$     3.7 &    12.88 $\pm$     1.86 & \nodata & \nodata & \nodata &  155 $\pm$   17 &  247 $\pm$   64 \\ 
                          Mrk493 &    62.69 $\pm$     1.31 &             $<$   23.47 &    27.09 $\pm$     1.53 &    28.42 $\pm$     4.76 &     9.59 $\pm$     1.27 & \nodata & \nodata &  280 $\pm$   39 &  267 $\pm$   46 &  280 $\pm$   38 \\ 
                          Mrk590 &    32.81 $\pm$     2.12 &             $<$   16.58 &    21.14 $\pm$     2.54 &    31.04 $\pm$     3.06 &     9.57 $\pm$     1.64 & \nodata & \nodata &  197 $\pm$   33 &  200 $\pm$   33 &  206 $\pm$   40 \\ 
                          Mrk704 &             $<$   10.05 &             $<$   61.44 &             $<$   50.82 &             $<$  151.67 &    51.08 $\pm$     3.04 & \nodata & \nodata & \nodata & \nodata &  249 $\pm$   37 \\ 
                          Mrk705 &    52.30 $\pm$     1.90 &             $<$   23.26 &    49.59 $\pm$     2.11 &    57.82 $\pm$     2.19 &    30.18 $\pm$     1.66 & \nodata & \nodata &  192 $\pm$   31 &  210 $\pm$   35 &  280 $\pm$   39 \\ 
                         NGC3227 &   708.7 $\pm$    25.2 &   220.9 $\pm$     5.5 &   725.8 $\pm$     6.2 &   668.0 $\pm$    17.6 &   231.2 $\pm$     7.5 & \nodata &  241 $\pm$   32 &  228 $\pm$   30 &  216 $\pm$   29 &  209 $\pm$   34 \\ 
                         NGC3516 &    74.01 $\pm$     2.29 &   128.9 $\pm$     4.4 &   164.4 $\pm$     2.9 &   458.4 $\pm$     5.5 &    67.57 $\pm$     2.18 & \nodata & \nodata &  198 $\pm$   30 &  229 $\pm$   28 & \nodata \\ 
                         NGC4051 &   172.6 $\pm$     4.9 &             $<$   47.59 &   162.0 $\pm$     2.9 &   347.6 $\pm$     7.1 &   107.7 $\pm$     3.9 & \nodata & \nodata &  172 $\pm$   29 &  295 $\pm$   27 & \nodata \\ 
                         NGC4235 &    34.85 $\pm$     1.62 &             $<$    6.03 &    33.29 $\pm$     1.77 &    33.60 $\pm$     3.04 &             $<$    5.28 & \nodata & \nodata &  214 $\pm$   34 &  286 $\pm$   43 & \nodata \\ 
                         NGC5548 &    83.08 $\pm$     3.32 &    42.90 $\pm$     2.55 &    81.76 $\pm$     2.77 &   124.7 $\pm$     8.6 &    31.50 $\pm$     2.16 & \nodata & \nodata &  177 $\pm$   30 & \nodata &  206 $\pm$   38 \\ 
                         NGC7469 &  1915 $\pm$    27 &    90.00 $\pm$     7.88 &   357.9 $\pm$     7.5 &   340.0 $\pm$    38.0 &   154.4 $\pm$    10.0 & \nodata &  158 $\pm$   12 &  196 $\pm$   29 &  196 $\pm$   45 &  240 $\pm$   44 \\ 
                      PG0026+129 &     2.56 $\pm$     0.33 &     4.94 $\pm$     0.43 &             $<$    7.72 &    20.16 $\pm$     4.01 &             $<$    5.07 & \nodata & \nodata & \nodata &  305 $\pm$   40 & \nodata \\ 
                      PG0804+761 &             $<$    4.44 &    15.53 $\pm$     2.28 &    20.89 $\pm$     1.70 &    20.74 $\pm$     2.87 &             $<$   11.13 & \nodata &  292 $\pm$  111 &  323 $\pm$   47 &  342 $\pm$   66 & \nodata \\ 
                      PG1119+120 &             $<$    4.00 &    20.63 $\pm$     2.55 &    28.41 $\pm$     2.01 &    60.00 $\pm$     2.37 &    16.16 $\pm$     2.33 & \nodata & \nodata & \nodata & \nodata &  193 $\pm$   47 \\ 
                      PG1229+204 &     5.84 $\pm$     0.59 &             $<$   20.90 &    12.05 $\pm$     1.37 &    26.74 $\pm$     3.55 &    11.48 $\pm$     1.13 & \nodata & \nodata & \nodata & \nodata &  218 $\pm$   50 \\ 
                      PG1351+640 &    21.92 $\pm$     0.99 &             $<$   10.26 &    29.39 $\pm$     1.73 &             $<$   12.38 &             $<$    3.83 & \nodata & \nodata &  200 $\pm$   43 & \nodata & \nodata \\ 
                      PG1411+442 &     4.54 $\pm$     0.79 &     7.79 $\pm$     1.00 &     9.24 $\pm$     0.62 &    13.99 $\pm$     2.31 &     6.49 $\pm$     0.58 & \nodata & \nodata & \nodata &  212 $\pm$   45 &  316 $\pm$   63 \\ 
                      PG1426+015 &    13.46 $\pm$     1.08 &    13.67 $\pm$     2.28 &    25.55 $\pm$     1.09 &    29.51 $\pm$     4.38 &     8.47 $\pm$     1.09 &  226 $\pm$   38 & \nodata &  258 $\pm$   36 &  292 $\pm$   57 &  289 $\pm$   57 \\ 
                      PG1440+356 &    42.72 $\pm$     0.96 &    15.55 $\pm$     1.70 &    38.57 $\pm$     1.26 &    50.22 $\pm$     2.37 &    14.26 $\pm$     0.86 &  193 $\pm$   34 &  211 $\pm$   39 &  253 $\pm$   35 &  253 $\pm$   38 &  304 $\pm$  107 \\ 
                      PG1613+658 &    37.66 $\pm$     1.23 &    13.25 $\pm$     1.01 &    30.90 $\pm$     1.02 &    66.49 $\pm$     3.15 &             $<$    7.92 &  226 $\pm$   30 &  269 $\pm$   33 &  289 $\pm$   33 &  225 $\pm$   29 & \nodata \\ 
                      PG2130+099 &    16.14 $\pm$     1.22 &    32.54 $\pm$     1.53 &    55.73 $\pm$     2.88 &   100.8 $\pm$     3.7 &    45.16 $\pm$     3.01 & \nodata & \nodata &  187 $\pm$   30 & \nodata &  287 $\pm$   45 \\ 
                      PG2349-014 &    15.95 $\pm$     0.71 &             $<$    6.36 &    20.51 $\pm$     1.01 &    34.73 $\pm$     3.82 &             $<$    6.83 &  273 $\pm$   48 & \nodata &  259 $\pm$   34 &  308 $\pm$   35 & \nodata \\ 
 		\hline
                         3C120 &    78.86 $\pm$     3.72 &   225.7 $\pm$     3.3 &   271.6 $\pm$     3.7 &  1129 $\pm$     7 &   163.8 $\pm$     2.7 & \nodata & \nodata & \nodata &  114 $\pm$    8 &  111 $\pm$   16 \\ 
                        Mrk509 &   118.0 $\pm$     4.1 &    73.19 $\pm$     3.92 &   153.0 $\pm$     3.4 &   180.0 $\pm$     5.1 &    54.04 $\pm$     2.63 &  173 $\pm$   35 & \nodata & \nodata &  165 $\pm$   30 & \nodata  \\ 
                   NGC3783 &   197.0 $\pm$     3.3 &   128.8 $\pm$     5.1 &   255.9 $\pm$     5.0 &   380.0 $\pm$     8.5 &   152.0 $\pm$     5.2 & \nodata & \nodata &  244 $\pm$   12 &  141 $\pm$   24 &  133 $\pm$   21 \\ 
                    NGC4151 &  1180 $\pm$    12 &  1130 $\pm$    22 &   350.0 $\pm$    13.9 &  2030 $\pm$    33 &   560.0 $\pm$    11.0 & \nodata &  207 $\pm$   16 &  155 $\pm$    7 &  169 $\pm$   12 &  139 $\pm$   25  \\

 \enddata
                         
\tablecomments{No data for the line fluxes indicate no spectral coverage at the observed-frame wavelengths of these lines. No data for the velocity dispersion of detected lines indicate that their signal to noise ratio was between 3 and 5, or that their profiles were unresolved.}
\end{deluxetable}
\end{centering}

\clearpage
\begin{centering}
\begin{deluxetable}{ccccccccccc}
\rotate
\tablecolumns{11}
\tabletypesize{\tiny}
\tablewidth{0pt}
\tablecaption{\label{tab:type2_ir} Fluxes and Velocity Dispersions of Ionized Gas Narrow Lines in type 2 AGNs.}
\tablehead{ 

\colhead{Galaxy} & \colhead{f$_{\rm [Ne~\sc{II}]} $} & \colhead{f$_{\rm[S~\sc{IV}]} $} & \colhead{f$_{\rm [Ne~\sc{III}]} $} & \colhead{f$_{\rm [O~\sc{IV}]} $} & \colhead{f$_{\rm [Ne~\sc{V}]} $} & \colhead{\sneii} & \colhead{\ssiv} & \colhead{\sneiii} & \colhead{\soiv} & \colhead{\snev} \\
\colhead{(-)} & \colhead{ (10$^{-18}$ W m$^{-2}$)} & \colhead{(10$^{-18}$ W m$^{-2}$)} & \colhead{(10$^{-18}$ W m$^{-2}$)} & \colhead{(10$^{-18}$ W m$^{-2}$)} & \colhead{(10$^{-18}$ W m$^{-2}$)} & \colhead{(km s$^{-1}$) } & \colhead{(km s$^{-1}$) } & \colhead{(km s$^{-1}$) } & \colhead{(km s$^{-1}$) }  & \colhead{(km s$^{-1}$)  } 
}

\startdata
          2MASXJ08035923+2345201 &             $<$    6.39 &             $<$    5.50 &    18.78 $\pm$     2.08 &    28.45 $\pm$     1.59 &             $<$    4.75 & \nodata & \nodata &  203 $\pm$   42 & \nodata & \nodata \\ 
          2MASXJ08244333+2959238 &    33.17 $\pm$     2.04 &             $<$   20.88 &    49.47 $\pm$     3.44 &    83.64 $\pm$     2.13 &    36.14 $\pm$     2.78 &  207 $\pm$   39 & \nodata & \nodata & \nodata &  264 $\pm$   30 \\ 
          2MASXJ10181928+3722419 &     8.67 $\pm$     1.53 &             $<$    6.35 &    14.69 $\pm$     1.87 &    28.81 $\pm$     1.60 &    16.59 $\pm$     2.80 & \nodata & \nodata & \nodata & \nodata &  200 $\pm$   32 \\ 
          2MASXJ12384342+0927362 &             $<$    8.33 &    20.09 $\pm$     2.25 &    27.85 $\pm$     2.15 &    58.83 $\pm$     2.38 &    12.94 $\pm$     2.35 & \nodata &  232 $\pm$   42 &  260 $\pm$   42 &  261 $\pm$   36 &  265 $\pm$   73 \\ 
          2MASXJ16164729+3716209 &             $<$    5.34 &    16.53 $\pm$     1.49 &    22.55 $\pm$     2.52 &    90.14 $\pm$     2.28 &    17.88 $\pm$     2.23 & \nodata & \nodata & \nodata & \nodata &  211 $\pm$   37 \\ 
                     CGCG218-007 &   102.6 $\pm$     3.1 &    35.72 $\pm$     2.64 &    86.07 $\pm$     2.69 &   166.7 $\pm$     2.3 &    46.59 $\pm$     1.98 & \nodata & \nodata & \nodata & \nodata &  215 $\pm$   33 \\ 
                     ESO103-G035 &   294.0 $\pm$     7.6 &   115.1 $\pm$    10.1 &   414.0 $\pm$     9.4 &   311.7 $\pm$    13.9 &   172.9 $\pm$    10.5 &  223 $\pm$   37 &  232 $\pm$   34 &  276 $\pm$   31 &  240 $\pm$   35 &  346 $\pm$   52 \\ 
                  IRAS05189-2524 &   191.8 $\pm$     8.9 &    65.30 $\pm$     6.26 &   186.1 $\pm$     7.1 &   261.2 $\pm$    37.2 &   152.6 $\pm$    12.0 &  253 $\pm$   42 &  335 $\pm$   55 &  392 $\pm$   41 &  444 $\pm$   63 &  404 $\pm$   43 \\ 
                  IRAS15001+1433 &    66.48 $\pm$     1.36 &     4.74 $\pm$     0.82 &    27.97 $\pm$     0.69 &    21.78 $\pm$     3.21 &    15.89 $\pm$     1.97 &  247 $\pm$   34 &  388 $\pm$   36 &  348 $\pm$   38 &  558 $\pm$   96 &  422 $\pm$   81 \\ 
                  IRAS18325-5926 &   367.9 $\pm$     7.8 &   114.9 $\pm$     8.2 &   436.0 $\pm$     5.3 &   390.7 $\pm$    12.2 &   262.6 $\pm$     9.6 &  181 $\pm$   31 &  265 $\pm$   39 &  367 $\pm$   32 &  300 $\pm$   34 &  404 $\pm$   64 \\ 
                  IRAS23060+0505 &    32.26 $\pm$     2.40 &    19.65 $\pm$     2.34 &    25.61 $\pm$     2.10 &    35.31 $\pm$     5.45 &    20.12 $\pm$     2.69 &  278 $\pm$   59 &  366 $\pm$  127 &  285 $\pm$   62 &  343 $\pm$   57 &  419 $\pm$   62 \\ 
                   MCG-03-34-064 &   514.9 $\pm$     8.2 &   476.1 $\pm$     8.3 &  1150 $\pm$    13 &  1053 $\pm$    22 &   615.6 $\pm$     8.2 &  344 $\pm$   34 &  298 $\pm$   33 &  361 $\pm$   28 &  273 $\pm$   30 &  435 $\pm$   36 \\ 
                         Mrk1066 &  1094 $\pm$    21 &   102.0 $\pm$     7.4 &   469.1 $\pm$     7.6 &   418.1 $\pm$    22.1 &    92.38 $\pm$     8.31 & \nodata & \nodata &  216 $\pm$   32 &  284 $\pm$   41 &  246 $\pm$   48 \\ 
                         Mrk1457 &    91.99 $\pm$     2.39 &             $<$   14.09 &    38.13 $\pm$     4.09 &    28.10 $\pm$     2.54 &    22.09 $\pm$     3.04 & \nodata & \nodata & \nodata &  228 $\pm$   43 &  310 $\pm$   60 \\ 
                          Mrk273 &   444.9 $\pm$     7.9 &   101.7 $\pm$     2.3 &   338.1 $\pm$     2.5 &   550.1 $\pm$    17.8 &   112.3 $\pm$     3.7 &  252 $\pm$   30 &  357 $\pm$   34 &  332 $\pm$   30 &  357 $\pm$   38 &  406 $\pm$   32 \\ 
                            Mrk3 &   979.8 $\pm$    10.2 &   592.7 $\pm$     6.2 &  1749 $\pm$    11 &  1964 $\pm$    24 &   632.5 $\pm$     7.5 &  276 $\pm$   29 &  322 $\pm$   29 &  314 $\pm$   29 &  263 $\pm$   28 &  327 $\pm$   31 \\ 
                         Mrk463E &   108.2 $\pm$     3.5 &   275.0 $\pm$     5.4 &   404.6 $\pm$     7.3 &   641.8 $\pm$    11.3 &   193.9 $\pm$     4.7 &  308 $\pm$   35 &  284 $\pm$   30 &  235 $\pm$   29 &  295 $\pm$   29 &  287 $\pm$   31 \\ 
                          Mrk609 &   213.5 $\pm$     6.2 &             $<$   22.18 &    56.71 $\pm$     2.53 &    82.43 $\pm$    25.69 &    38.69 $\pm$     4.28 & \nodata & \nodata &  202 $\pm$   36 &  \nodata &  488 $\pm$   60 \\ 
                          Mrk622 &    97.29 $\pm$    13.80 &             $<$    8.61 &    49.29 $\pm$     2.82 &             $<$   37.42 &             $<$   38.06 & \nodata & \nodata &  333 $\pm$   34 & \nodata & \nodata \\ 
                         NGC1068 &  4988 $\pm$   183 &  6199 $\pm$   336 & 13781 $\pm$   246 & 20406 $\pm$   473 &  8974 $\pm$   221 &  359 $\pm$   52 &  414 $\pm$   44 &  383 $\pm$   31 &  364 $\pm$   48 &  363 $\pm$   39 \\ 
                         NGC1275 &   461.5 $\pm$     8.0 &             $<$   16.26 &   223.7 $\pm$     5.6 &             $<$   84.37 &             $<$   11.65 &  299 $\pm$   31 & \nodata &  260 $\pm$   31 & \nodata & \nodata \\ 
                         NGC2622 &    62.02 $\pm$     2.63 &             $<$   18.46 &    80.41 $\pm$     2.93 &    89.29 $\pm$     3.58 &    24.73 $\pm$     2.08 & \nodata & \nodata &  213 $\pm$   33 & \nodata &  289 $\pm$   39 \\ 
                         NGC2623 &   550.1 $\pm$     9.7 &    11.20 $\pm$     1.36 &   147.2 $\pm$     1.7 &   117.4 $\pm$    18.2 &    37.55 $\pm$     2.97 & \nodata & \nodata &  213 $\pm$   31 &  357 $\pm$   52 &  222 $\pm$   39 \\ 
                         NGC2639 &    86.23 $\pm$     2.66 &             $<$    6.90 &    48.88 $\pm$     1.95 &             $<$   23.82 &             $<$    7.51 &  346 $\pm$   43 & \nodata &  336 $\pm$   49 & \nodata & \nodata \\ 
                         NGC3079 &  1310 $\pm$    60 &             $<$   14.61 &   238.3 $\pm$     3.3 &             $<$   94.19 &             $<$   38.79 &  198 $\pm$   33 & \nodata &  285 $\pm$   29 & \nodata & \nodata \\ 
                         NGC4258 &   123.5 $\pm$     7.3 &             $<$   15.61 &    70.61 $\pm$     5.61 &    79.40 $\pm$     8.03 &             $<$   13.67 &  193 $\pm$   42 & \nodata &  191 $\pm$   36 &  204 $\pm$   53 & \nodata \\ 
                         NGC4507 &   307.8 $\pm$     6.9 &    93.08 $\pm$     6.84 &   287.8 $\pm$     6.3 &   354.1 $\pm$    11.6 &   125.7 $\pm$     6.4 & \nodata & \nodata &  220 $\pm$   35 &  318 $\pm$   42 &  345 $\pm$   81 \\ 
                         NGC5256 &   160.6 $\pm$     3.1 &    27.06 $\pm$     1.56 &    94.40 $\pm$     1.45 &   569.8 $\pm$    10.3 &    20.54 $\pm$     1.37 &  227 $\pm$   29 &  239 $\pm$   37 &  227 $\pm$   30 &  191 $\pm$   28 &  202 $\pm$   40 \\ 
                         NGC5506 &   850.6 $\pm$    14.3 &   735.4 $\pm$    15.6 &  1537 $\pm$    11 &  2262 $\pm$    40 &   568.0 $\pm$    10.6 & \nodata &  239 $\pm$   33 &  207 $\pm$   28 & \nodata &  208 $\pm$   32 \\ 
                         NGC5728 &   366.2 $\pm$     7.7 &   317.2 $\pm$     8.2 &   536.5 $\pm$     6.7 &  1155 $\pm$    12 &   217.1 $\pm$     3.4 & \nodata &  272 $\pm$   41 &  263 $\pm$   29 &  306 $\pm$   30 &  314 $\pm$   32 \\ 
                         NGC5929 &   104.8 $\pm$     2.2 &    12.41 $\pm$     1.98 &    95.07 $\pm$     1.40 &    43.67 $\pm$     6.18 &             $<$    6.09 & \nodata & \nodata &  210 $\pm$   28 & \nodata & \nodata \\ 
                         NGC6240 &  1872 $\pm$    29 &    38.7 $\pm$     2.9 &   599.9 $\pm$     4.2 &   374.0 $\pm$    35.9 &    90.24 $\pm$     9.02 &  362 $\pm$   37 &  279 $\pm$   52 &  341 $\pm$   29 & \nodata & \nodata \\ 
                         NGC7172 &   334.6 $\pm$     8.1 &    55.74 $\pm$     4.43 &   168.1 $\pm$     4.0 &   384.4 $\pm$     5.3 &    90.23 $\pm$     3.49 &  256 $\pm$   30 & \nodata &  204 $\pm$   30 & \nodata & \nodata \\ 
                         NGC7674 &   215.7 $\pm$     5.2 &   152.6 $\pm$     5.9 &   346.0 $\pm$     6.5 &   442.9 $\pm$    12.1 &   184.4 $\pm$     7.0 & \nodata &  267 $\pm$   34 &  253 $\pm$   32 &  242 $\pm$   32 &  267 $\pm$   39 \\ 
                     SBS1133+572 &   108.3 $\pm$     2.9 &    42.01 $\pm$     3.62 &   105.3 $\pm$     3.8 &   168.7 $\pm$     2.8 &    69.89 $\pm$     2.87 & \nodata &  248 $\pm$   40 & \nodata & \nodata & \nodata \\ 
                        UGC02608 &   566.5 $\pm$     9.4 &   292.8 $\pm$     9.5 &   706.3 $\pm$     6.1 &  1378 $\pm$    17 &   312.9 $\pm$     5.1 & \nodata &  213 $\pm$   31 & \nodata & \nodata & \nodata \\ 
                         UGC5101 &   379.7 $\pm$     9.0 &             $<$    7.34 &   146.2 $\pm$     2.1 &    57.18 $\pm$     9.66 &    34.19 $\pm$     3.73 &  254 $\pm$   40 & \nodata &  350 $\pm$   34 &  285 $\pm$   41 &  385 $\pm$   40 \\ 
                         \hline
                          Centaurus A &  2210 $\pm$    45 &   140.0 $\pm$     9.6 &  1400 $\pm$    12 &   980.0 $\pm$    42.2 &   200.0 $\pm$     8.5 &  220 $\pm$   31 & \nodata &  134 $\pm$   21 &  127 $\pm$    9 &  132 $\pm$   21 \\ 
                            Circinus &  4536 $\pm$   145 &  1270 $\pm$    58 &  4000 $\pm$    90 &  6793 $\pm$   420 &  2180 $\pm$   101 & \nodata &   91 $\pm$    6 &   86 $\pm$    4 &  113 $\pm$    3 &   90 $\pm$    8  \\ 
\enddata

\tablecomments{Same as in Table~\ref{tab:type1_ir}}
\end{deluxetable}
\end{centering}

\clearpage
\begin{centering}
\begin{deluxetable}{cccc}
\tablecolumns{4}
\tabletypesize{\tiny}
\tablewidth{0pt}
\tablecaption{\label{tab:mbh2}  Estimates of \mbh\ in type-2 AGN.}
\tablehead{ 

\colhead{Galaxy} & \colhead{\mbh ($\sigma_{*}^{4.24}$ )}   & \colhead{\mbh ($\sigma_{\rm NLR}^{4.24}$ )}  & \colhead{\mbh\ (plane equation)}  \\
\colhead{(-)} & \colhead{(10$^{7}$ \msun)} & \colhead{(10$^{7}$ \msun)}  & \colhead{(10$^{7}$ \msun)}
}

\startdata         
          2MASXJ08035923+2345201 &   2.1 &   9.3 &   4.4 \\
          2MASXJ08244333+2959238 &   0.9 &  17 &   7.5 \\
          2MASXJ10181928+3722419 &   0.7 &   5.3 &   7.3 \\
          2MASXJ12384342+0927362 &  23 &  19 &  16 \\
          2MASXJ16164729+3716209 &  23 &   6.6 &  22 \\
                     CGCG218-007 &   3.8 &   7.2 &   7.3 \\
                     ESO103-G035 & \nodata &  29 &   8.4 \\
                  IRAS05189-2524 &   2.7 & 132 &  29 \\
                  IRAS15001+1433 &  70 & 225 &  41 \\
                  IRAS18325-5926 & \nodata &  70 &  17 \\
                  IRAS23060+0505 & \nodata &  75 &  45 \\
                   MCG-03-34-064 & \nodata &  76 &  22 \\
                         Mrk1066 & \nodata &  20 &   7.9 \\
                         Mrk1457 &   3.2 &  24 &  10 \\
                          Mrk273 &  31 &  82 &  28 \\
                            Mrk3 & \nodata &  40 &  19 \\
                         Mrk463E &   5.5 &  26 &  36 \\
                          Mrk609 &   3.4 & 120 &  14 \\
                          Mrk622 &   2.5 &  76 &   9.0 \\
                         NGC1275 &  32 &  27 &  11 \\
                         NGC2622 & \nodata &  18 &   8.2 \\
                         NGC2623 &   4.1 &  37 &   7.5 \\
                         NGC2639 &  10 &  80 &   4.7 \\
                         NGC3079 &   3.9 &  40 &   3.0 \\
                         NGC4507 & \nodata &  41 &   8.4 \\
                         NGC5256 &   9.9 &   9.2 &   8.1 \\
                         NGC5506 & \nodata &   8.8 &   6.0 \\
                         NGC5728 & \nodata &  32 &   8.7 \\
                         NGC5929 &   1.7 &  11 &   3.0 \\
                         NGC6240 &  23 &  51 &  17 \\
                         NGC7172 & \nodata &   9.5 &   4.0 \\
                         NGC7674 & \nodata &  19 &  16 \\
                     SBS1133+572 &  15 &  12 &  11 \\
                        UGC02608 & \nodata &   6.0 &  11 \\
                         UGC5101 &  10 &  71 &  17  \\
\enddata

\tablecomments{
The second column corresponds to the result of the stellar \msigma\ relation.
The third column presents the \mbh\ estimate using the \msigma\ relation for the NLR,
as found in Figure~\ref{fig:msigma}. The fourth column uses the plane equation presented
in Figure~\ref{fig:plane}. For the NLR gas-based estimates, we present the average value 
using all resolved lines. Formal, statistical \mbh\ uncertainties are of a factor of 0.5 dex. 
For individual sources the uncertainty can be higher, typically within an order of magnitude.
}
\end{deluxetable}
\end{centering}

\end{onecolumn} }

\end{document}